\documentclass[usenatbib,useAMS,usegraphicx]{mn2e}

\usepackage{times}

\usepackage{xspace}
\usepackage{amsmath}
\usepackage[rightcaption]{sidecap}
\usepackage{comment}
\usepackage{mathtools}

\output\expandafter{\the\output\floatfix}

\makeatletter
\def\floatfix{%
\expandafter\ifx\csname r@x@one\endcsname\relax
\else
\ifnum\c@page=\numexpr\expandafter\expandafter\expandafter
              \@secondoftwo\csname r@x@one\endcsname-1\relax
\aftergroup\figone
\fi
\fi}
\makeatother

\providecommand{\eg    }{e.g.\xspace}%
\providecommand{\ie    }{i.e.\xspace}
\providecommand{\mrk   }{Mrk~421\xspace}%
\providecommand{\xray  }{X-ray\xspace}%
\providecommand{\xrays }{X-rays\xspace}%
\providecommand{\gray  }{$\gamma$-ray\xspace}%
\providecommand{\grays }{$\gamma$-rays\xspace}%
\providecommand{\fermi }{\textit{Fermi}\xspace}%
\providecommand{\rxte}{\textit{RossiXTE}\xspace}%
\providecommand{\wise}{\textit{WISE}\xspace}%
\providecommand{\asca}{\textit{ASCA}\xspace}%
\providecommand{\maxi}{\textit{MAXI}\xspace}

\usepackage{colordvi}
\newcommand\be{\begin{equation}}
\newcommand\ee{\end{equation}}
\newcommand\bal{\begin{align}}
\newcommand\eal{\end{align}}

\title[Diffusion-Acceleration in Jets: Stochastic Variation ]
{Particle diffusion and localized acceleration in inhomogeneous AGN jets -
Part II: stochastic variation}
\author[X. Chen et al.]{%
\parbox{\textwidth}{Xuhui~Chen$^{1,2}$\thanks{chenxuhui.phys@gmail.com},
  Martin~Pohl$^{1,2}$, 
  Markus~B\"ottcher$^{3,4}$,
  Shan~Gao$^{2,1}$
}\vspace{0.1cm}\\
$^1$ Institute of Physics and Astronomy, University of Potsdam, 14476 Potsdam-Golm, Germany\\
$^2$ DESY, Platanenallee 6, 15738 Zeuthen, Germany\\
$^3$ Centre for Space Research, North-West University, Potchefstroom 2520, South Africa\\
$^4$ Astrophysical Institute, Department of Physics and Astronomy, Ohio University, Athens, OH 45701, USA\\
      }

\begin{document}

\maketitle

\label{firstpage}

\begin{abstract}
We study the stochastic variation of blazar emission under a 2-D spatially resolved
leptonic jet model we previously developed.
Random events of particle acceleration and injection 
in small zones within the emission region are assumed to be responsible for
flux variations.
In addition to producing spectral energy distributions that describe
the observed flux of \mrk,
we further analyze the timing properties of the simulated light curves,
such as the power spectral density (PSD) at different bands, flux-flux correlations, as well as the
cross-correlation function between
\xrays and TeV \grays.
We find spectral breaks in the PSD at a timescale comparable to 
the dominant characteristic time scale in the system,
which is usually the pre-defined decay time scale of an acceleration event. Cooling imposes a delay, and 
so PSDs taken at lower energy bands in each emission component (synchrotron or inverse Compton) generally break at longer timescales.
The flux-flux correlation between \xrays and TeV \grays can be either 
quadratic or linear, depending on whether or not there are large variation of
the injection into the particle acceleration process.
When the relationship is quadratic, the TeV flares lag the \xray flares,
and the optical \& GeV flares are large enough to be comparable to 
the ones in \xray.
When the relationship is linear, the lags are insignificant,
and the optical \& GeV flares are small.

\end{abstract}
\section{Introduction}
\label{intro}

Blazars are a special group of Active Galactic Nuclei (AGN)
whose observed radiation is dominated by relativistic jets.
They show erratic
variability at almost all electromagnetic wavelengths 
\citep{ulrich_maraschi_urry:1997:review}.
The most extreme variations are observed in \grays, where amplitudes
can be as large as a factor of 50 
\citep[\eg][]{veritas_2014:mrk421_14year:54.1}, and the doubling
time of the flux can be as short as 3 minutes 
\citep{hess_2007:pks2155_exceptional_flare}.
Variability in different wavebands is usually correlated
\citep[\eg][]{fossati_2008:xray_tev,chatterjee_2008:psd:689.79} 
with notable exceptions \citep{krawczynski_etal:2004:1es1959,
hess_2009:2155_lowstate:696.150}.
Because most of the blazar emission comes from non-thermal radiation of
plasma in the relativistic jets that moves in a direction highly
aligned with our line of sight, strong relativistic beaming effects
alter the temporal and spectral features of the emission 
\citep{urry_padovani:1995:review}, contributing to the violent variability of blazars.
For example, it can shorten the timescales of the variations,
and make them appear much faster than
the intrinsic variation timescales in the jets' comoving frame. 
But even with this considered,
significant variability on the timescale of several minutes is
still not readily explained for jets with sizes
presumably larger than the Schwarzschild radii of the black holes 
\citep{hess_2007:pks2155_exceptional_flare}.
In fact, most of those extreme temporal features observed in blazars 
are still yet to be understood.

The rich information carried within time-series data may also provide vital
clues for the discrimination of currently degenarate emission models,
namely the leptonic and hadronic models of high-energy emission 
\citep{boettcher_2013.leptonic_hadronic.768.54}.
The former can be further classified into synchrotron self-Compton (SSC) models
\citep{maraschi_1992:3c279_ssc:397.5}
and external Compton (EC) models
\citep{dermer_1992:disc, ghisellini_1996:blr:280.67, sikora_etal:2009:torus:704.38}, 
while the latter consists of proton synchrotron 
\citep{muecke_2001:proton_synchrotron},
p-p pion production
\citep{pohl_2000:blastwave:354.395}, 
and p-$\gamma$ pion production
\citep{mannheim_1992:pgamma:253.21, mannheim:1993:proton_blazar} models.
Furthermore, detailed time-dependent modeling of the blazar jets can
help constrain the physical conditions such as emission location, 
magnetic field, size, and even composition of the jets 
\citep{petropoulou_2014:tdep_hadronic_2155:571.83, diltz_2015:tdep_hadronic:802.133, sokolov_marscher_mchardy:2004:SSC,zacharias_2013:sync_lightcurve_ssc.777.109}.

Time-dependent modeling of blazars usually refers to
well defined, 'clean' flares
\citep{saito_2015:tdep_model_1510, chen_etal:2011:multizone_code_mrk421,
zhang_2015:3c279:804.58}, 
because only in those cases can one easily identify
a single physical event that may possibly reproduce the observation.
However, there is no guarantee that such clean flares are representative of
the general behavior of blazar variability. In fact, clean flares with 
predictable raising and decaying phases are relatively rare 
\citep{nalewajko_2012:brightest_flares}. Most of the
light curve may be better described as a succession of flares that
overlap each other, if one insists on describing light curves with flares
\citep{hess_2007:pks2155_exceptional_flare,
fermi_2010:3c279:463.919}.
An alternative approach is to continuously change one or more parameters 
in the model so that the resulting light curves match an entire observation
campaign in at least one waveband
\citep{krawczynski_coppi_aharonian:2002:timedep}.
This approach keeps the modeling close to observation and provides a good
way to probe the underlying physical connection between the variations at 
different wavebands.
However, it does not offer any physical reasoning why the physical parameters
change in the way the fitting process requires them, as
the timing information contained in the primarily fitted wavelength is largely 
left unexplored.

The noise-like appearance of blazar light curves triggered interests
in their power spectral density (PSD), which can reveal the distribution
of flux variability on different timescales 
\citep[\eg][]{chatterjee_2008:psd:689.79, abdo_etal:2010:light_curves_and_variability}.
Studies show that blazar PSDs typically exhibit a featureless power-law
spectrum close to an $1/f^2$ red-noise profile. However, the range of
timescales covered in those studies is usually limited, because it is 
naturally difficult to collect time-series data over vastly different timescales.
Therefore it is entirely possible that there are PSD spectral breaks
that can not yet be robustly identified from the currently available data.
For example, \citet{kataoka_2001:asca_421_psd:560.659} analyzed the 
0.5-10 keV \xray variability of \mrk,
and the results suggested a break of the PSD indices
around $10^{-5}$ Hz, close to the longest time scale accessible with their data.
A more recent analysis of 3-10 keV \xray variability
on longer time scales
\citep{isobe_2015:maxi_421_psd:798.27} confirms a relatively
soft PSD index of 1.6, although that is much harder than predicted by
\citet{kataoka_2001:asca_421_psd:560.659}.

The randomness in the blazar behavior prompted attempts to treat the
blazar physics as stochastic processes, either using the internal shock
models \citep{spada_etal:2001:internal_shocks, guetta_etal:2004:internal_shocks, rachen_2010:discontinuous:1006.5364} or turbulence  \citep{marscher_2014:temz_model.780.87} to
explain blazar variability.
With these approaches, it is practically impossible to match observed and simulated
light curves, because the variation of the simulated 
flux at any particular time is random. However, the fluctuations obey
certain probability distributions, so that over a long time, the
light curves would produce predictable PSDs. Those PSDs, as simulation
products, can be directly compared with the PSDs obtained from past or future 
observations. Additionally, the correlation between different
wavebands remains very informative of the inherent physical processes.
The traditional fitting of blazar spectral energy distributions (SEDs) 
will also be available just as in
steady-state or time-dependent approaches. Combining
all this information, one can take advantage of the large amount of
multiwavelength data available
and place strong constraints on blazar models.

In \citet{chen_2015:diffusion:447.530} (Paper I hereafter), we have built 
an inhomogeneous jet-emission 
model. As an initial step towards a more comprehensive model, we studied 
the spectral properties of the system as it relaxes to its steady state.
In this work we will expand that model by introducing stochastic processes,
building time-dependent SEDs and simulating light curves to be compared
with observations. The aim is to identify distinctive timing features, 
that can be used to explain observed variability and
guide future observation or data analysis.

In this paper, all quantities in the comoving frame of the jet are primed, while quantities
in the observer's frame are unprimed. Notice that this is contrary to the
convention in Paper I, because in that work most of the discussions are focused
on particle processes in the comoving frame, while in this work the focus
is shifted to observational consequences.

\section{Stochastic Processes and PSDs}
\label{sec:psd}

The PSD of evenly spaced time-series data can be calculated conveniently.
For the ease of comparison with observations, we use the formula of 
\citet{uttley_2002:psd:332.231} and \citet{chatterjee_2008:psd:689.79} 
to calculate PSDs from observational light curves of blazars. First, the average flux is subtracted off all flux points in a light curve, yielding
\begin{equation}
\label{eq:norm}
  \begin{split}
f(t_i) & =F(t_i)-\frac{1}{N}\,\sum\limits_{j=1}^N F(t_j) \\
       & =F(t_i)-\mu\ .
  \end{split}
\end{equation}
The power spectral density then is 
\begin{equation}
	P(\nu) =\frac{2T}{\mu^2N^2}|F_N(\nu)|^2\ ,
	\label{eq:psd}
\end{equation}
where  
\begin{equation}
	|F_N(\nu)|^2=
	|\sum\limits_{i=1}^N f(t_i)\mathrm{exp}(i2\pi\nu t_i)|^2 .
	\label{eq:fft}
\end{equation}

Here, T is the total
duration of the light curves considered,
N is the total number of data points,
and $\mu$ is the average flux subtracted.
Eq.\,\ref{eq:fft} calculates the square of the modulus of the discrete Fourier transform for the light curves at linearly evenly sampled frequencies $i/T$,
where $i$ are integers between 1 and $N/2$. Eq.\,\ref{eq:psd} gives the fractional rms-squared normalization to the final PSD.

With the fractional rms-squared normalization used in Eq.\,\ref{eq:psd},
the PSDs have the property that the integral of the PSDs give the total 
fractional rms variability of the light curve.
They give a direct view of how variable the emission is at different time scales. 
Of special interest are the slopes of the PSDs, 
which may reveal key properties of the underlying stochastic process.

In the context of discrete time series, white noise results from perhaps the most simple stochastic process. It consists of a sequence of
random variables that are independent of each other. 
The PSD of white noise has slope zero, \ie the power per unit frequency is uniformly distributed
on all time scales.

If the signal from one time step persist to the next time step, and the 
random variable is allowed to accumulate, then the resulting time series becomes red noise. This stochastic process
is a random walk process.
The continous-time limit of a Gaussian random walk in one-dimension is called
the Wiener process, which also produces red noise. The PSD of red noise
has slope -2, with more variabiliy power per unit frequency distributed at large time scales.

Some further modification of the Wiener process can also allow for breaks 
in the slope of PSDs. One such modified process is the Ornstein-Uhlenbeck (O-U)
process. It allows the accumulated signal in the Wiener process to exponentially decrease
with time, on time scale $\tau$, \eg imposed by particles escape/cooling or turbulence decay. This decrease means that
at the extreme of large time scales, the earlier signal almost vanishes 
(acceleration events decay), and the signal approaches white noise; at the extreme of small time scales, the
signal does not have enough time to decrease (acceleration events accumulate), and the process approaches a
Wiener process. The PSD slope changes from 0 at large time scale to -2 at small
time scale. The slope breaks approximately at frequency $1/(2\pi\tau)$.
An analytical expression for the PSD of an O-U process exists 
\citep{kelly_2009:OUprocess:698.895}:
\begin{equation}
	P(\nu)=\frac{2\sigma^2\tau^2}{1+(2\pi\tau \nu)^2}\ .
	\label{eq:ou_psd}
\end{equation}
Here $\sigma$ is related to the strength of the O-U process, and is only 
relevant to the normalization of the PSD. For a mathematical description of the O-U process, we refer the readers to \citet{gillespie_1996:brownian:64.225}.

\begin{figure}
\includegraphics[width=0.98\linewidth]{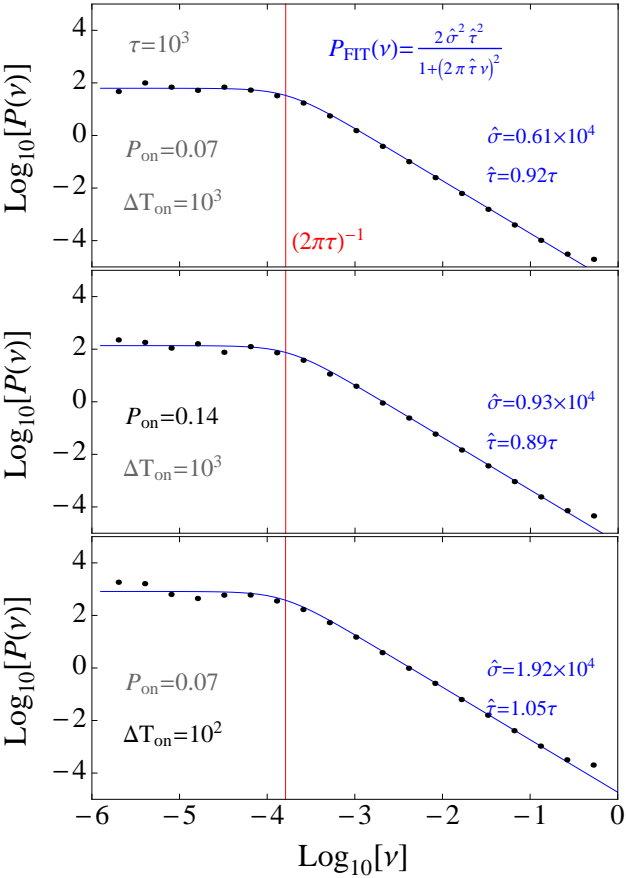}
\caption{PSD of a stochastic process similar to the O-U process. 
The black dots are simulation results, while the black legends are input parameters.
The blue lines are fitting to the black dots with the equation in blue.
The blue parameters are results of the fitting.
The vertical red line shows the frequency where we expect
to see a PSD break in an O-U process. Compared to the top panel, the middle panel has a different $\Delta T_\mathrm{on}$, while the bottom panel has a different $P_\mathrm{on}$.}
\label{fig:math_psd}
\end{figure}

As a mathematical model that motivates the implementation of stochastic 
variations in our jet model, 
we simulate a slightly modified version
of the O-U process where instead of Gaussian distribution, the increment 
follows the Bernoulli distribution, \ie it takes value of 1 with probability $P_\mathrm{on}$, and takes 0 otherwise. The increment is only active on 
certain time steps with separation of
$\Delta T_\mathrm{on}$, which is, for example, $10^3\Delta t$. Here 
$\Delta t$ is the time step, as well as the unit of time in these simulations.
There is no extra input, \ie the increment remains 0, 
on other time steps. The simulations last 6000 time steps each.

The simulated time sequence is used to produce PSDs according to eq. 
\ref{eq:norm}-\ref{eq:fft}. The resulting $\sim 10^6$ PSD points are grouped into 19 bins, where the bins have the same width on the logarithmic scale in the frequency axis (x-axis of the plot). The points shown in Figure \ref{fig:math_psd} represent the average values from each bin. 
Three plots are shown in a column
to illustrate the effects of changing the increments separation, 
$\Delta T_\mathrm{on}$, and the success probability, $P_\mathrm{on}$.

We can see that in all these plots, above the frequency defined by $1/(2\pi\tau)$, the PSDs are 
compatible with red noise, whereas below the break frequency they are 
consistent with white noise. This is the expected characteristic of the 
O-U process. We fit eq. \ref{eq:ou_psd} to the simulated PSDs by allowing
both parameters $\sigma$ and $\tau$ to vary. In practice, 
the logarithm of $P(\nu)$ is used for the 
fitting to avoid dominance of the high-power points at low frequency.

Also shown in Figure \ref{fig:math_psd} is that the fitted $\hat{\tau}$ is always 
very close to the actual $\tau$ in the model. 
This shows that the break in the PSD is solely determined by the decay time scale, 
while not being affected by the parameters
$\Delta T_\mathrm{on}$ or $P_\mathrm{on}$, as shown in the middle and 
bottom panels of Fig.\ref{fig:math_psd}.

\section{Model Setup}
\label{sec:model}

The model used in this work is built upon the spatially resolved treatment of particle acceleration and diffusion in jets that is presented in Paper I,
where the discussion focused on the steady-state spectrum.
The particle transport equation used in the model reads as 
\begin{equation}
\label{eq:FPeq}
\begin{split}
 \frac{\partial n' (\gamma',\mathbf{r'},t')}{\partial t'} & =
-\frac{\partial}{\partial \gamma'}\bigg[\dot{\gamma'}(\gamma',\mathbf{r'},t')n'(\gamma',\mathbf{r'},t')\bigg] \\
 &  +\frac{\partial}{\partial \gamma'}\bigg[D'(\gamma',\mathbf{r'},t')
 \frac{\partial n'(\gamma',\mathbf{r'},t')}{\partial \gamma'} \bigg] \\
& +\mathbf{\nabla}\cdot\bigg[D'_x(\gamma') \nabla n'(\gamma',\mathbf{r'},t')\bigg] + Q'(\gamma',\mathbf{r'},t') ,
\end{split}
\end{equation}
where $n'(\gamma',\mathbf{r'},t')$ is the differential number density of particles,
$\dot{\gamma'}(\gamma',\mathbf{r'},t')$ is the acceleration rate (
including the synchrotron and SSC cooling, which are negative),
$D'(\gamma',\mathbf{r'},t')$ is the momentum diffusion coefficient, 
$D'_x(\gamma')$ is the spatial diffusion coefficient,
and $Q'(\gamma',\mathbf{r'},t')$ is the particle injection rate.
Further explanation of this equation, along with a schematic figure for the geometry
can be found in Paper I. 
A more detailed description of the Monte Carlo/Fokker-Planck (MCFP) code
used for building the model can be found in \citet{chen_etal:2011:multizone_code_mrk421}.
In the current work we adopt the Dirichlet boundary conditions 
($n'_\mathrm{boundary}=0$), 
\ie the spatial diffusion will lead to 
particle escape through the outer boundary of the simulation box. 
The Monte-Carlo photons in the simulation output 
are binned according to the angle of their traveling direction 
to the jet axis, $\theta'$. In order to get sufficient number of Monte Carlo photons for the bin
that has Doppler factor $\delta \sim \Gamma$, we sample a finite range of $\theta'$ around $90^\circ$ in the jet frame ($-0.12\le \mathrm{cos}\theta'\le0.12$ here). But the relativistic beaming is performed using the average angle (cos$\theta'=0$) to avoid
any spread in the Doppler factor for the beaming, which can cause significant spread of the flux in time
when the simulation becomes relatively long.
The cylindrical simulation box is divided into axisymmetrical ring-like cells
in radial and vertical directions. The number of cells is nz$\times$nr=20$\times$15=300. The magnetic field is assumed to be homogeneous and disordered, 
so that the spatial diffusion is also isotropic. 
The radiation mechanism is assumed to be leptonic, with current discussions
limited to SSC models, even though our MCFP code also allows for the study of EC
models \citep{chen_2012:1510_ec:424.789}.
The simulated results (\S\,\ref{sec:results}) including SEDs, light curves and flux-flux correlations are plotted
together with the observations of \mrk, 
the stereotypical high-energy-peak blazar
that we try to compare our simulations with.
This matching to \mrk also influences our discussions, many of
which, especially the energy bands, may appear to be specific to \mrk.
However, most of the results are indeed applicable to other blazars,
especially other high synchrotron peak blazars.
The key parameters used in the
benchmark case (\S\,\ref{sec:injacc}) are listed in Table\,\ref{tab:par}.

\begin{table}
\tabcolsep 4pt
\centering
\caption{The parameters used for the benchmark case in \S \ref{sec:injacc}. The observation angle in the observer's frame is always $1/\Gamma$ so that
the Doppler factor $\delta$ is equal to the bulk Lorentz factor $\Gamma$.
The volume height $Z'=4R'/3$ in all cases. 
Here $n'_\mathrm{e}$ is the time average of the particle density of the entire region for $t'/\Gamma>100$~ks, while $t'_\mathrm{acc}$ and $Q'_\mathrm{inj}$ represent the fastest acceleration time and the associated injection rate in a single cell during the simulation.}
\label{tab:par}       
\begin{tabular}{*{9}{c}}
\hline
$B'$ & $Z'$ & $\delta$ & $D'_\mathrm{x}$ &
$n_\mathrm{e}'$ & $t'_\mathrm{acc}$ & $Q'_\mathrm{inj}$ & $\tau'_\mathrm{decay}$\\\hline
0.27 & $10^{16}$ & 33 & 3.75 &
1.53 & 0.18 & $0.020$ & 2 \\
Gauss & cm & - & $\mathrm{cm}^2\mathrm{s}^{-1}$ & $\mathrm{cm}^{-3}$ & $Z'/c$ & $\mathrm{cm}^{-2}\mathrm{s}^{-1}$ & $Z'/c$ \\
\hline
\end{tabular}
\end{table}

In order to achieve a stochastic
increase of localized acceleration, which we use to describe the turbulent
nature of the plasma, 
we set the acceleration rate in each cell  in a way similar to the stochastic process we simulated
in \S \ref{sec:psd}.
The specific implementations of the O-U-process-like acceleration in our jet-model is as follows:
In most parts of the emission region, the particles are assumed to only diffuse
spatially and cool radiatively, \ie there is no acceleration.
This assumption is broken in a spinal region along the axis of
the axisymmetrical jet (where the radius $r'<10^{15}$cm),
where second-order Fermi acceleration with acceleration rate 
$\dot\gamma'_{D}(\gamma',\mathbf{r'},t')=\gamma'/t'_\mathrm{acc}$ is at work. Initially, the acceleration time, $t'_\mathrm{acc}$, is very long.
Within this region, for every time step ($\delta t'=\delta z'/c=1.7\times10^{4}\,\mathrm{s}$ here), 
each sub-region with 2x2 cells may experience with a probability of $P_\mathrm{acc}$ (=0.07 here)
an increase in the
acceleration rate. The increment in acceleration rate is
\be
\Delta \dot\gamma'_{D}= \gamma'\,\frac{c}{2.2\,Z'}=\gamma'\,\frac{1}{7.3\cdot 10^5\ \mathrm{s}}\ .
\label{rate}\ee
These increase can happen repeatedly on the same cells.
Sub-regions can overlap with each other in the z direction, so that each 
cell belongs to two sub-regions and has two chances of
receiving additional acceleration every time step.
On the other hand, the acceleration rates also decrease exponentially on the time scale $\tau'_\mathrm{decay}$ all the time, everywhere.
The stochastic increase of the acceleration represents
additional turbulence caused by, \eg, magnetic reconnection, or other plasma processes that are inherently stochastic.
The deterministic decrease on the other hand represents the dissipation
of these magnetic turbulence through reconnection.

Most of the specific parameter values used in this section are tuned 
so that the simulated and observed SED and light curves look reasonably similar.
However, as demonstrated in the previous section, the choice of the
parameters like time-step and $P_\mathrm{acc}$ does not affect key features
such as the power-law breaks in the PSDs.
 
One important 
complication in our model is that here the random variable is the 
acceleration rate, rather than the radiative flux that is measured directly.
Because of this, several time scales besides the acceleration decay time scale 
become potentially relevant in the problem. Those include the radiative-cooling
time scale and light-crossing time scale.
We will further compare our simulation
results with the O-U process, and discuss the relevance of different time 
scales in \S\,\ref{sec:results} \& \ref{sec:discussion}.

\begin{figure}
\includegraphics[width=0.99\linewidth]{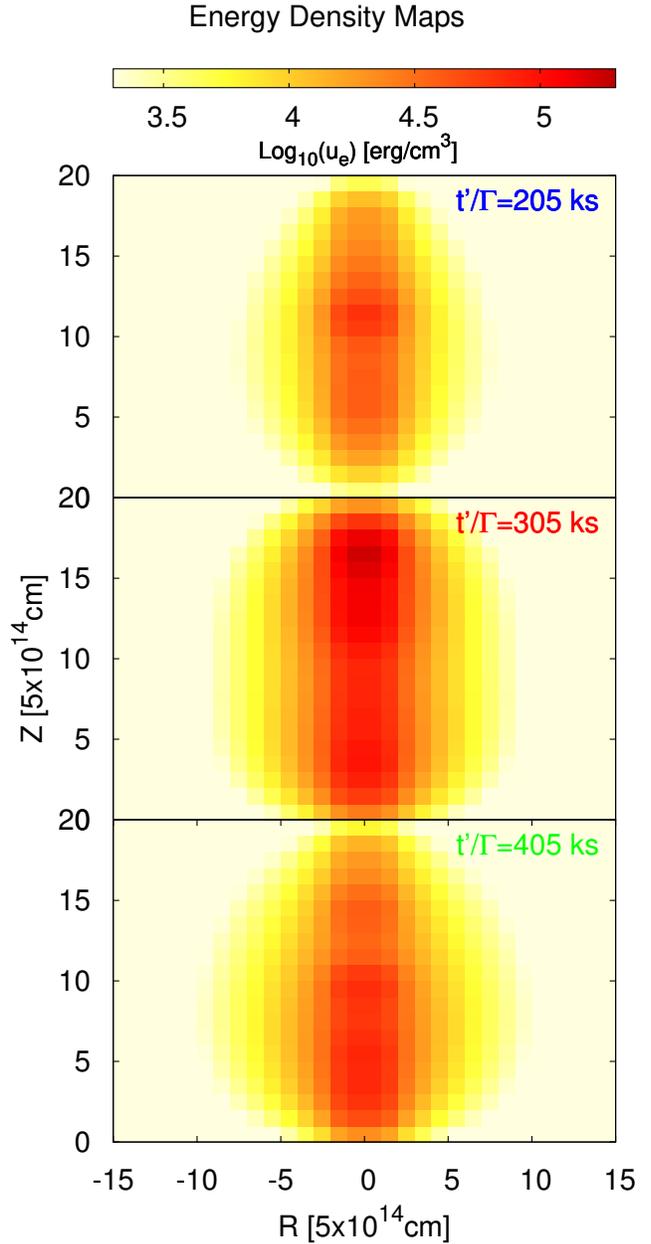}

\caption{Three sample electron energy density maps at different time epochs
  from the benchmark simulation case described in \S\,\ref{sec:injacc} (Case 1).
	$t'/\Gamma$ are shown as rough estimates for their corresponding time
	in the observer's frame, because proper relativistic 
	beaming is not defined for electrons.
}
\label{fig:emap}
\end{figure}

In the acceleration regions,
electrons are also injected at relatively low, 
but still highly relativistic energy.
Instead of following a Gaussian distribution as assumed in Paper I,
these electrons follow
the Maxwell-J\"uttner distribution \citep{wienke_1975:max_juentter:15.151, juettner_1911:max_juettner:339.856} expressed as a function of Lorentz factor $\gamma'$:
\begin{equation}
  f(\gamma') = \frac{\gamma'^2\beta'}{\Theta' K_2(1/\Theta')}e^{-\gamma'/\Theta'},
\end{equation}
where $K_2$ is a modified Bessel function of the second kind,
$\Theta'$ is associated to the temperature $T$ as $\Theta'=kT'/mc^2$,
$k$ is Boltzmann's constant, $m$ is the rest mass of the particle (electron here), 
$\beta'=v'/c$, and c is the speed of light. $\Theta'$ is chosen to equal the bulk 
Lorentz factor of the jet $\Gamma$, because the injected electrons are 
postulated as a result of particles in the interstellar medium being
trapped and isotropized by the magnetic turbulence which also causes
the particle acceleration in the jet. This is similar to the isotropization
of particles of the interstellar medium through blast waves as 
proposed by \citet{pohl_2000:blastwave:354.395}.

\begin{figure}
\includegraphics[width=0.99\linewidth]{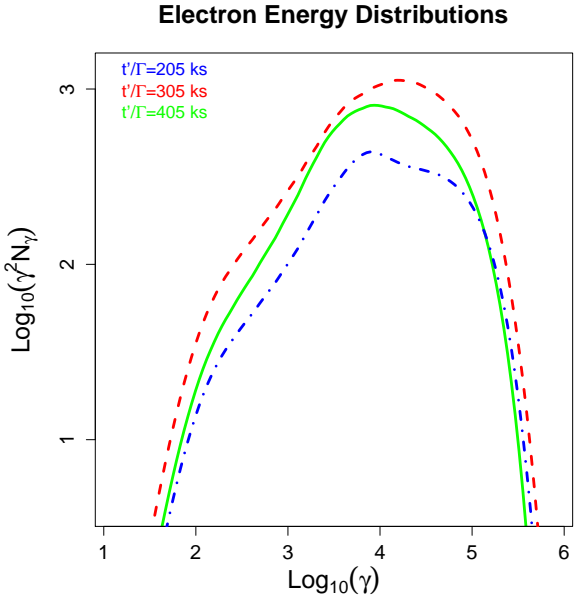}

\caption{Three sample electron energy distributions 
  from the benchmark case (Case 1).
  The time sequence is blue dot-dashed line, red dashed line, then green solid line.
	$t'/\Gamma$ are shown as rough estimates for their corresponding time
	in the observer's frame.
}
\label{fig:eed}
\end{figure}

An illustrative example of the 2-D energy-density maps of electrons in the simulation 
(Case 1) is plotted in Fig.\,\ref{fig:emap}.
The corresponding electron energy distributions averaged over the entire 
emission region are plotted in Fig.\,\ref{fig:eed}.
Both figures use quantities in the co-moving jet frame. To be noted from the figures are the asymmetry in the spatial distribution of electrons and the spectral impact of the stochasticity in acceleration efficiency.

Microphysical processes that provide an enhanced level of turbulence, and hence stronger stochastic acceleration, often go along with preacceleration of particles that subsequently undergo Fermi-type acceleration. To avoid specifying the origin of fast stochastic acceleration, we investigate two representative scenarios.
In the benchmark case, shown in \S\,\ref{sec:injacc}, the rate of particle injection at low energy, 
non-zero only in the acceleration region,
is exponentially coupled to the acceleration rate according to
\begin{equation}
  Q'_\mathrm{inj} = Q'_{0} \mathrm{exp}(t'_0/t'_\mathrm{acc}) ,
  \label{eq:exp_inj}
\end{equation}
where $Q'_0=10^{-7}\ \mathrm{cm}^{-2}\, \mathrm{s}^{-1}$ and $t'_0=2.16\, Z'/c$.
These values are tuned so that the 
changes in both acceleration and injection
affect the evolution of the light curves.
The motivation for the exponential coupling used here is:
  Suppose the injection is linearly associated with the high-energy tail of a background particle population, which falls off exponentially; and suppose
  the pre-acceleration of the background particles is linearly associated
with the localized acceleration. The injection is then exponentially associated with the peak energy of the background particles, and so is the acceleration rate.
Table\,\ref{tab:par} lists the fastest
injection rate we observed in the simulation.

The connection between injection and acceleration is changed in the second case (\S\,\ref{sec:cpick}, representing the alternative scenario), where injection is absent where the acceleration is slow ($t'_\mathrm{acc}\gg Z'/c$). When the acceleration rate is reasonably strong ($t'_\mathrm{acc}\sim Z'/c$ or $t'_\mathrm{acc} < Z'/c$), the injection rate is set to a constant ($Q'_\mathrm{inj} =4\times10^{-3}\ \mathrm{cm}^{-2}\,\mathrm{s}^{-1}$).
Effectively, this leads to a constant particle density of $1.49\ \mathrm{cm}^{-3}$. 

Under the same physical scenario as the one in the benchmark case, 
we investigate a third case, as outlined in \S\,\ref{sec:l80}.
This case explores the effects of the acceleration
decay rate, by changing $\tau'_\mathrm{decay}$ from $2Z/c$ to $4Z/c$. To
ensure the maximum acceleration rate achieved is more or less unchanged, we accordingly reduce in amplitude the stochastic increments of the acceleration.

Most simulations (except the duplicate simulation for Fig.\,
\ref{fig:421_combined}) in this work
use the same sequence of random numbers for the stochastic change of 
the acceleration rate
(but not for the Monte Carlo radiative transfer). 
This is evident from the similar light curves seen in all cases. 
This makes the comparison between different cases easier, and leaves the
intended change of condition in the system as the most likely cause for any differences.

\section{Results}
\label{sec:results}

\begin{figure*}
\includegraphics[width=0.49\linewidth]{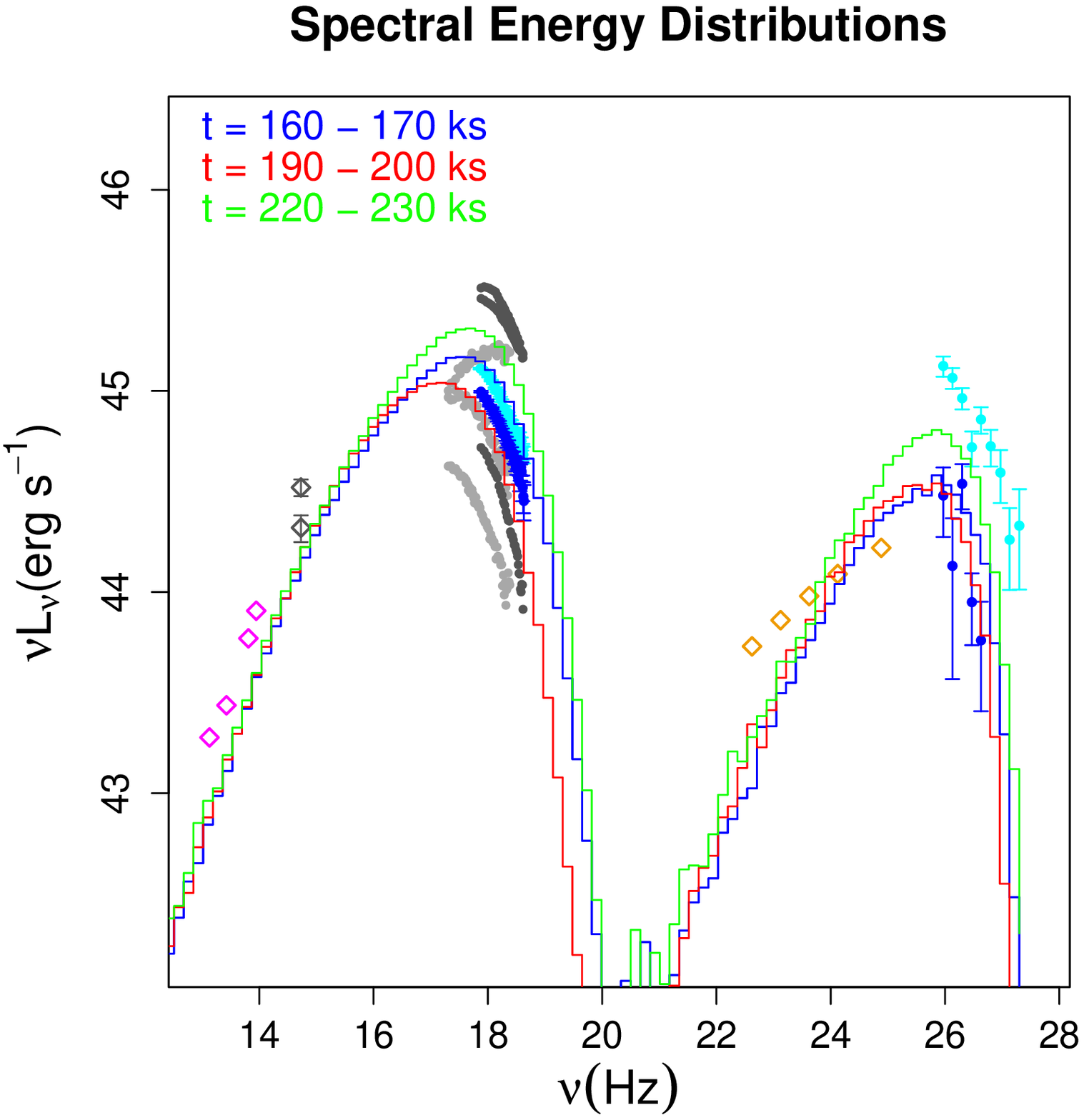}
\includegraphics[width=0.49\linewidth]{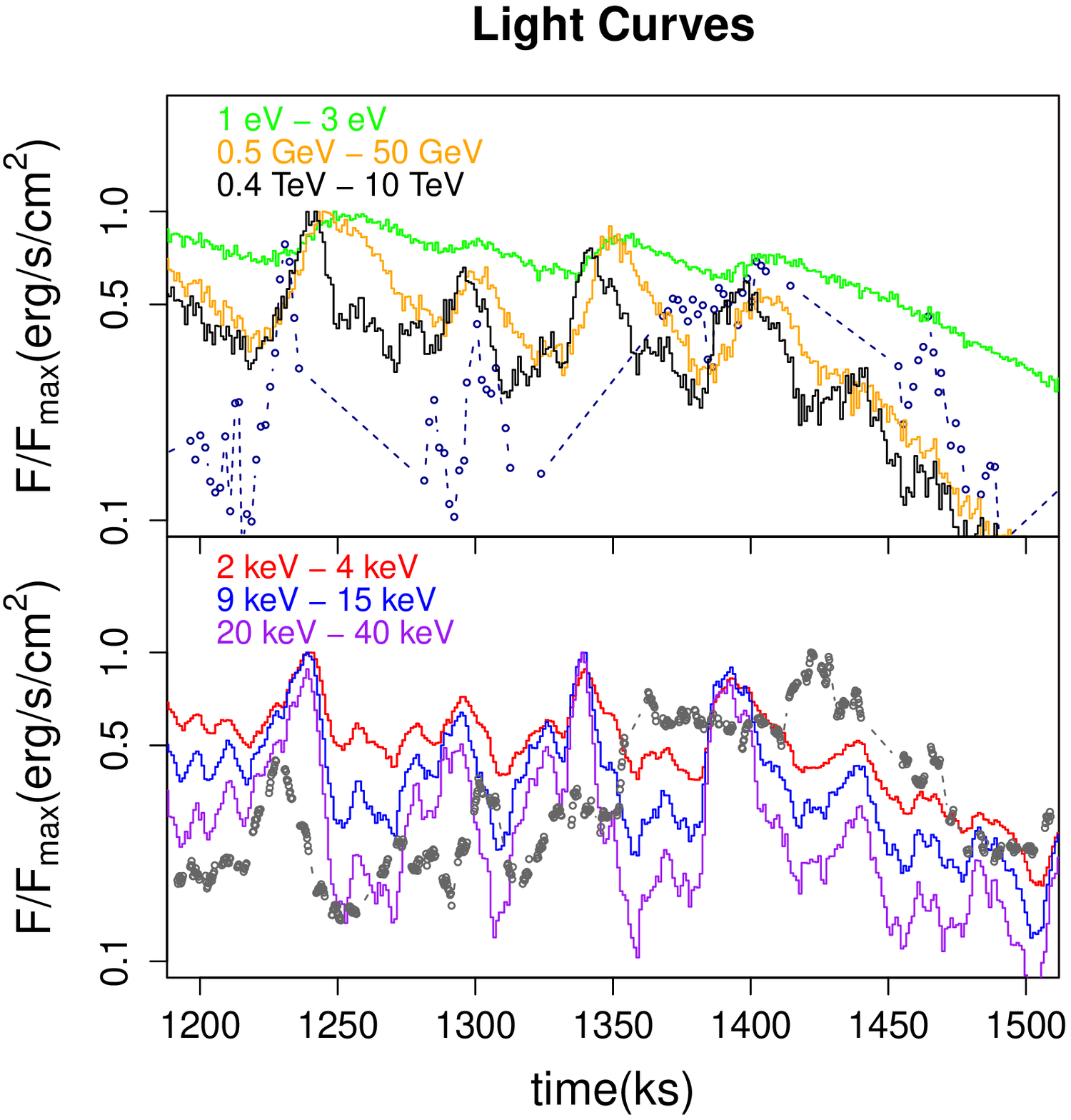}
\caption{Left: Simulated SED (histogram) at three different time epochs for our benchmark case (Case 1).
The data points include a \gray flare of 
March 19, 2001, observed by Whipple, and the simultaneous \xray data 
from \rxte (pre-flare in blue, flare peak in cyan). \xray and optical 
high/low states data taken by \rxte and Mt Hopkins 48-inch telescope 
in 2001 (dark gray), and \xray high/low states data from 
BeppoSAX in 1998 and 2000 (light gray) are also shown 
\citep{fossati_2008:xray_tev}. 
In other 
bands there are \wise infrared data from May 21, 2010 in magenta,
and \fermi 4-year average \gray data 
\citep{fermi_2015:4year_catalog:218.23} in yellow.
Right: Samples of normalized simulated light curves in 6 different frequency bands. Observational light curves are plotted as connected dots.
They include simultaneous 2-4 keV \rxte X-ray (grey) and TeV Whipple/HEGRA \gray (dark blue) data. The time stamp 100\ ks in the figure corresponds to MJD 59186.9 or March 18, 2001.
}
\label{fig:421_300}
\end{figure*}

\begin{figure*}
	\includegraphics[width=0.49\linewidth]{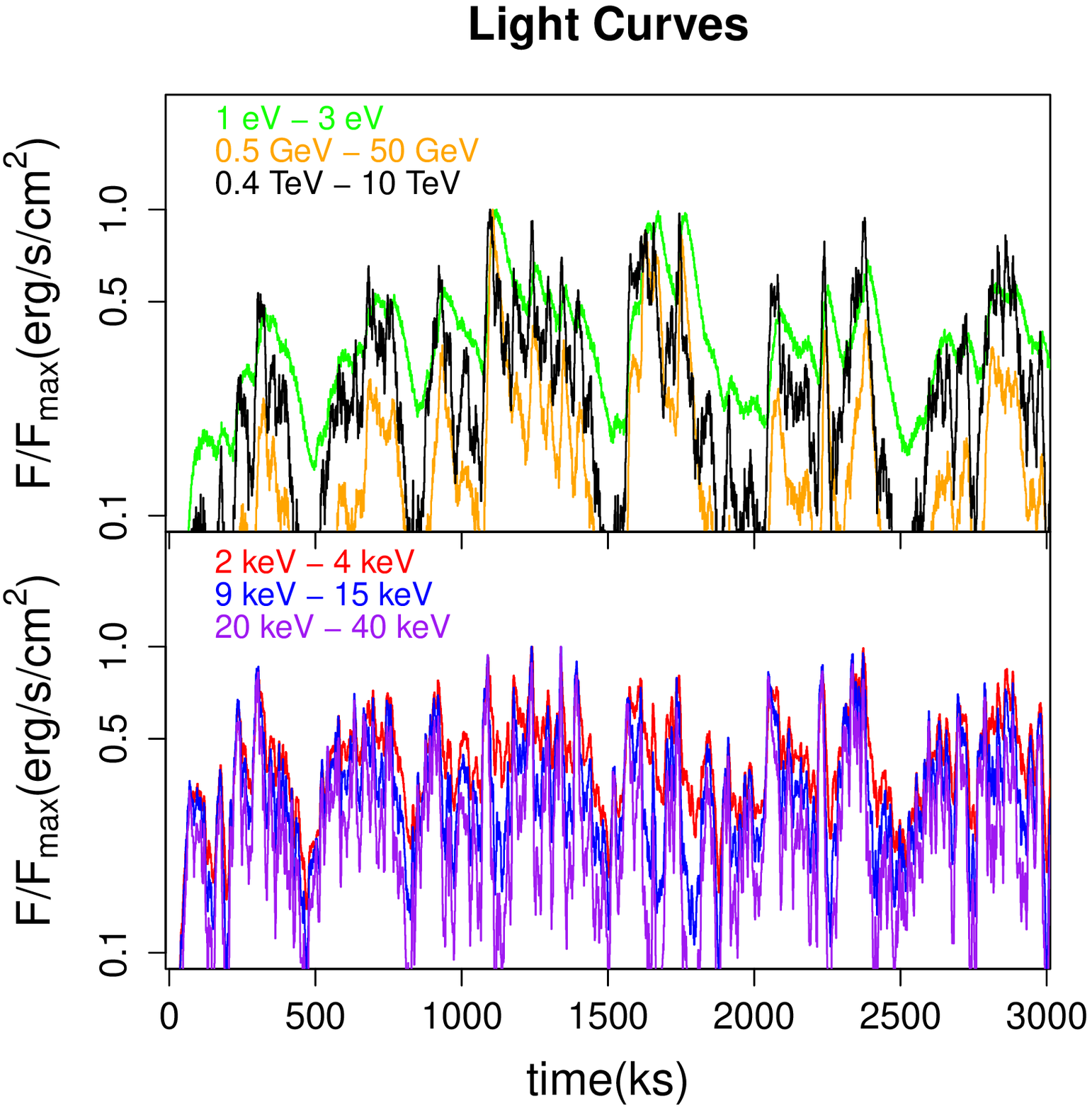}
	\includegraphics[width=0.49\linewidth]{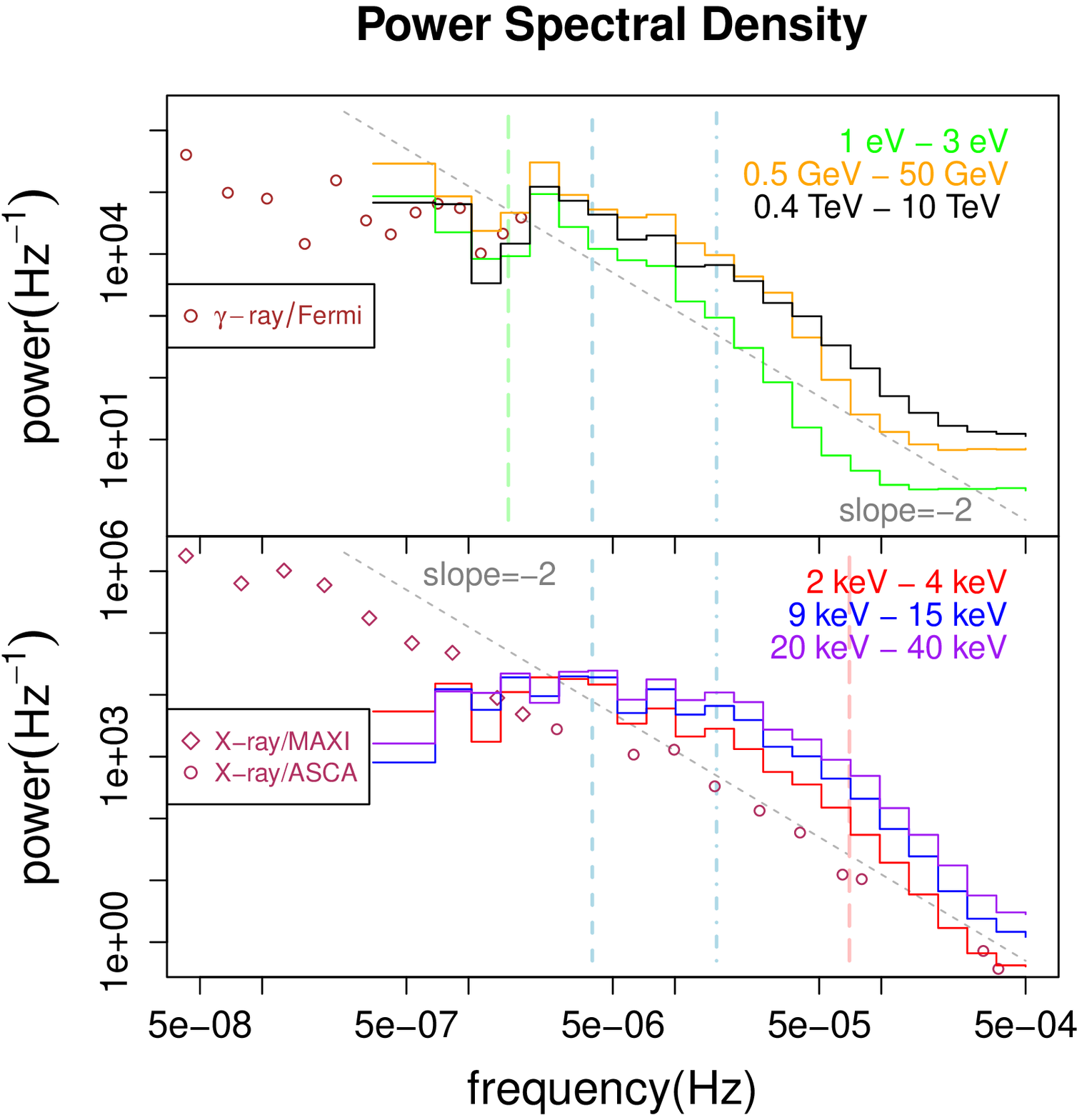}
	\caption{Left: Full light curves for the entire simulation Case 1.
Right: PSD based on the light curves excluding the first 100\ ks. The vertical lines indicate various timescales in the system (times $2\pi$), namely acceleration decay as blue dashed line, light crossing as blue dot-dashed line, and a red/green long dashed lines representing
synchrotron cooling for electrons that emit synchrotron radiation primarily in 2-keV \xray/ 1-eV optical band. 
The gray dashed line indicates a power-law with index of -2.
We also plot the observational PSDs of \mrk. They include \xray data from \maxi (squares) and \asca (circle) as maroon points in the lower panel \citep{isobe_2015:maxi_421_psd:798.27} and GeV \gray from \fermi as brown points in the upper panel.}
	\label{fig:421_full}
\end{figure*}

\begin{figure*}
	\includegraphics[width=0.49\linewidth]{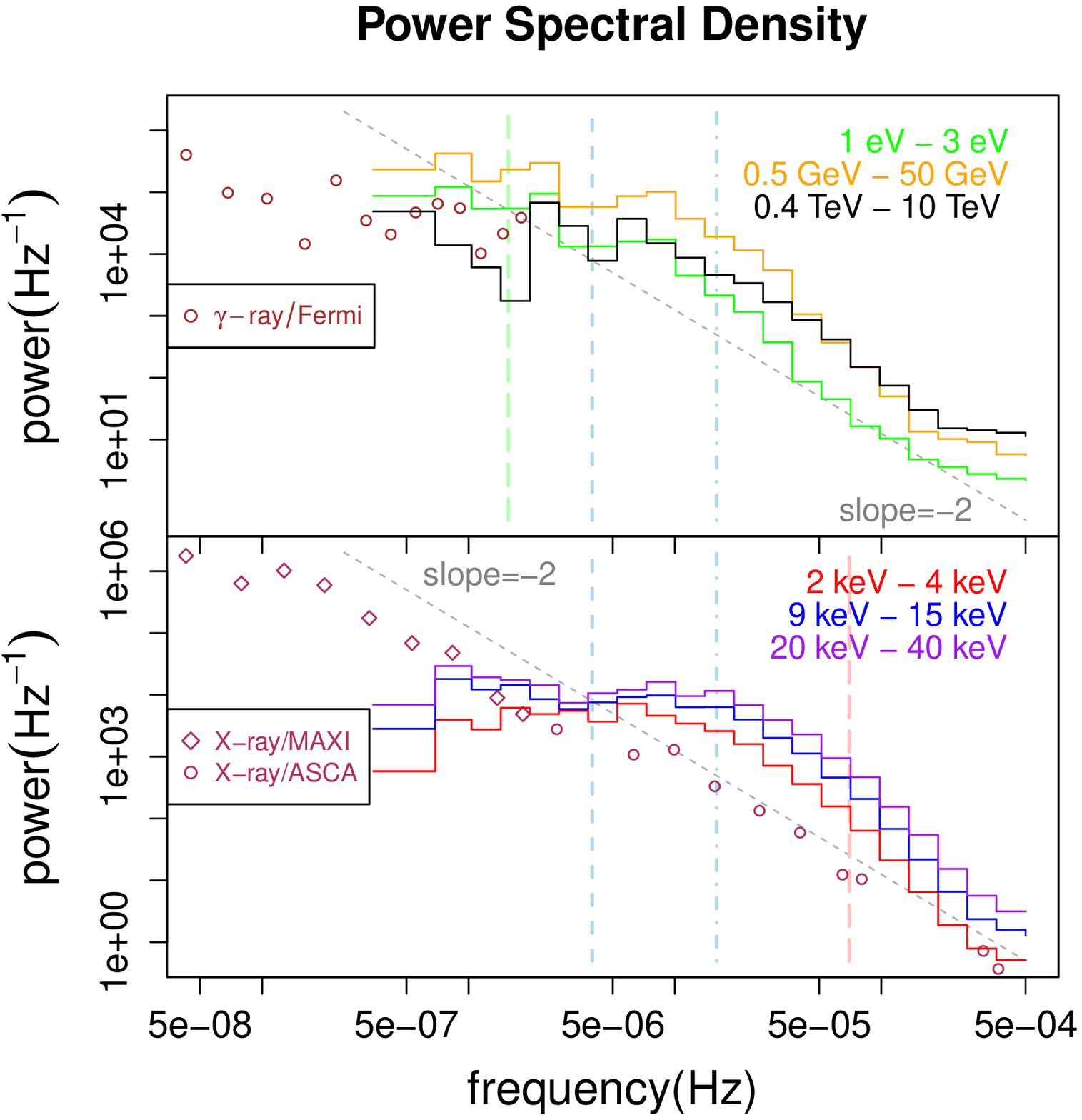}
	\includegraphics[width=0.49\linewidth]{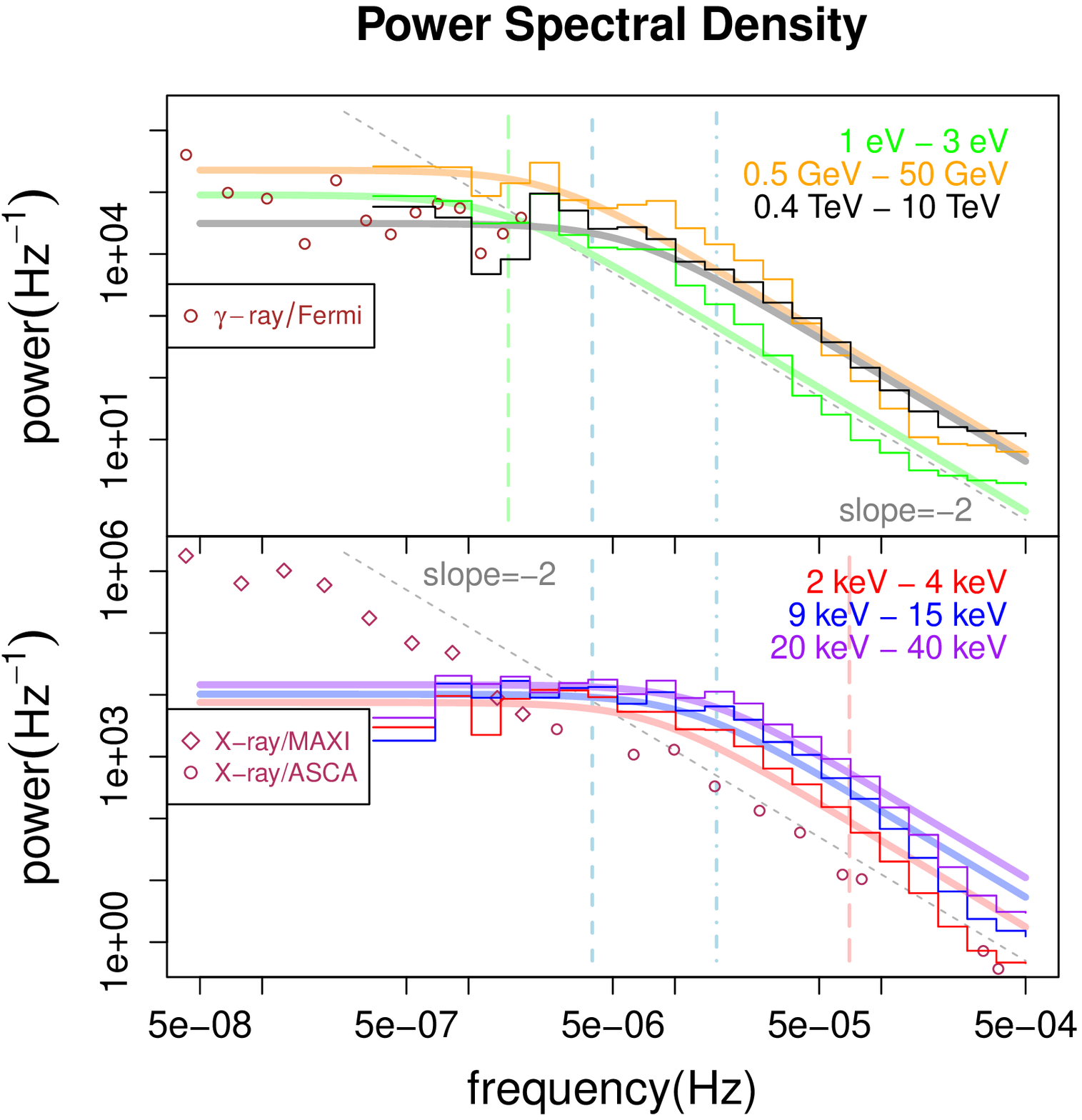}
	\caption{Left: Power spectral density for a simulation almost the same as Case 1, except a 
	different realization of the random numbers. Right: Average PSD of 
	the two simulations. Lines that show the O-U process fits to the PSD 
	are plotted with matching colors. The straight comparison
	lines are the same as those 
	in Fig.\,\ref{fig:421_full}.}
	\label{fig:421_combined}
\end{figure*}


\begin{figure*}
	\includegraphics[width=0.49\linewidth]{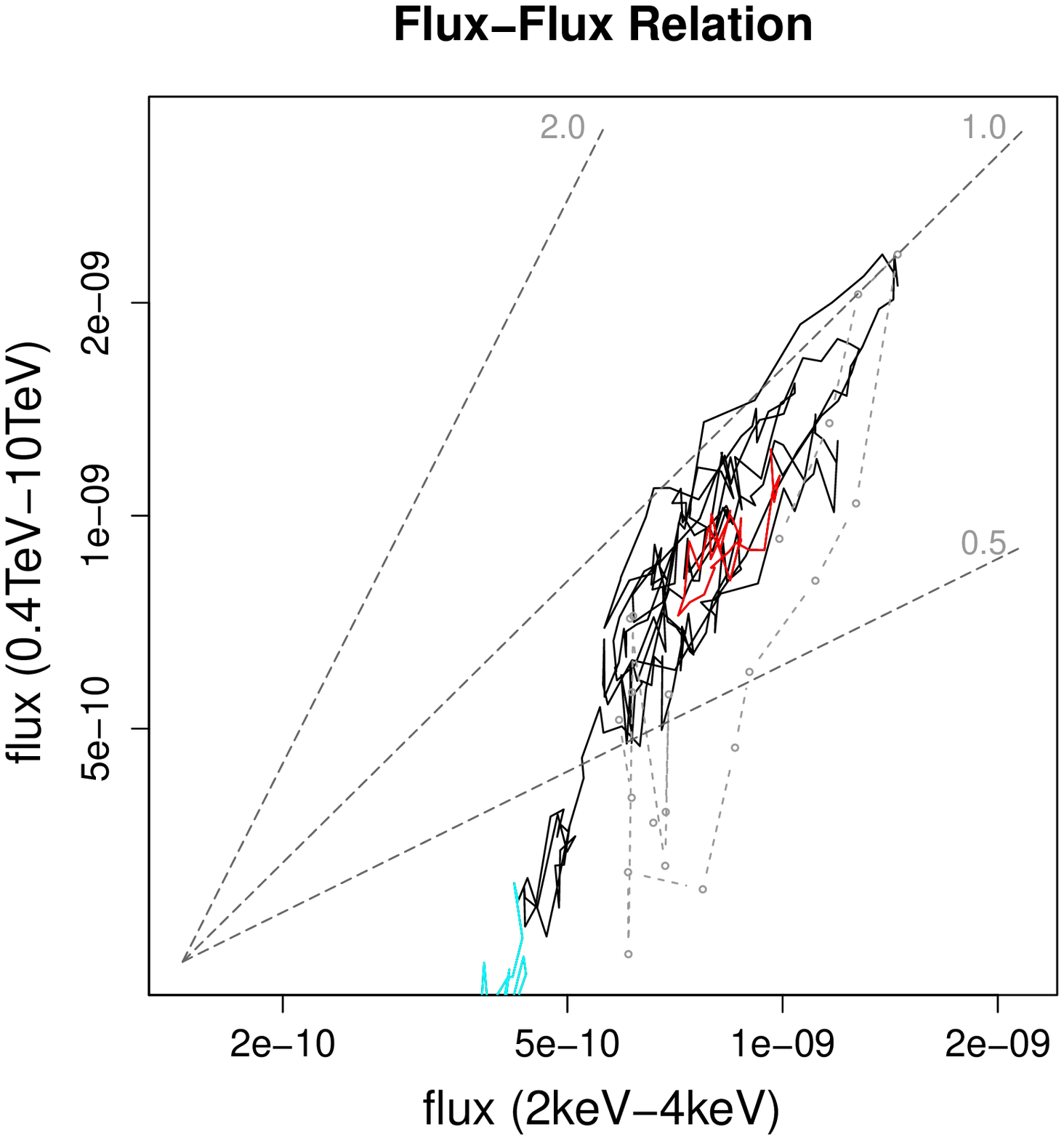}
	\includegraphics[width=0.49\linewidth]{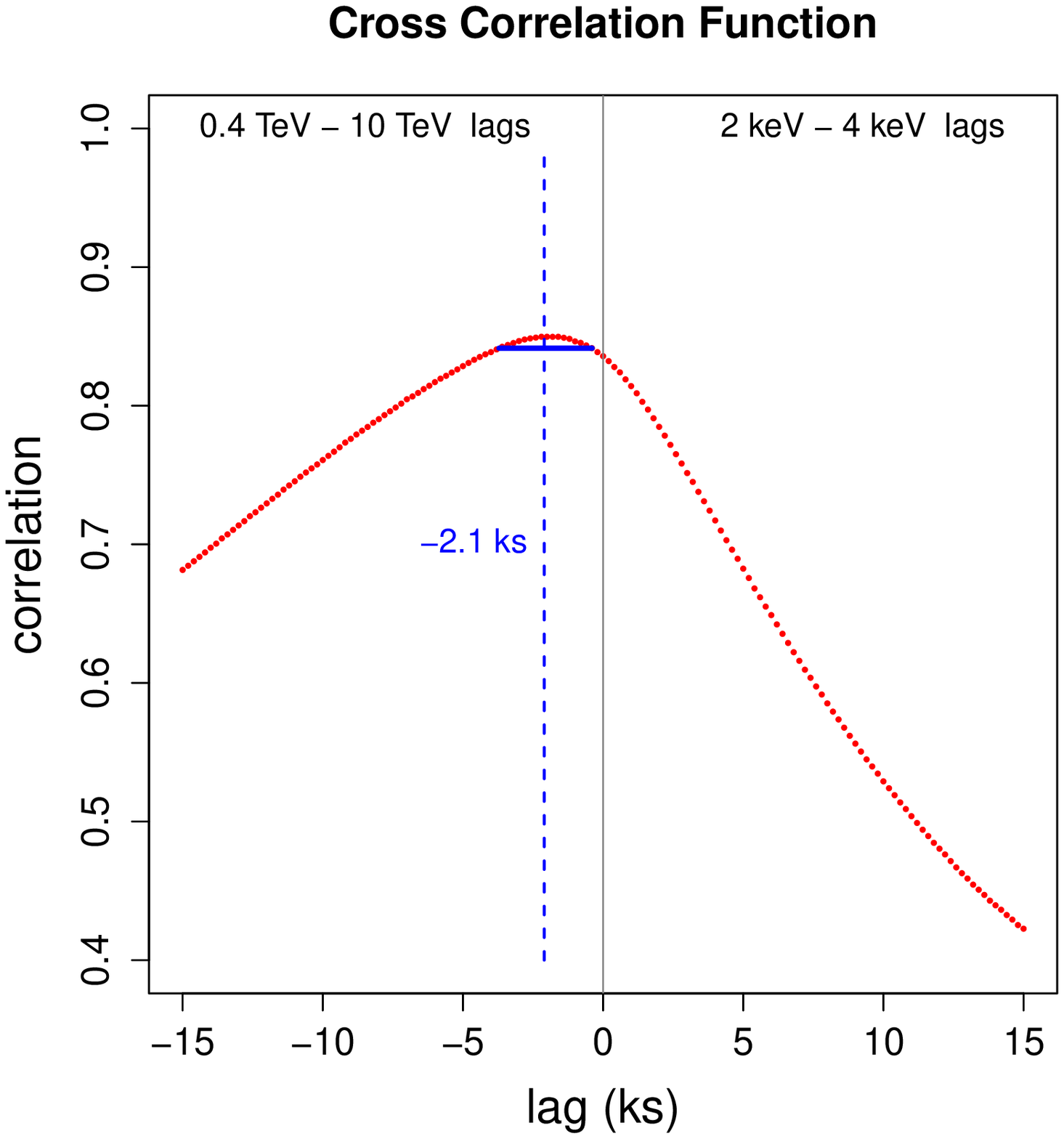}
	\caption{Correlation analysis for Case 1. Left: Flux-flux correlation between the \xray and \gray bands during 
	the time period
	1200ks-1500ks. The first 30\ ks are plotted in red, and the last 30\ ks are 
	plotted in cyan. The grey connected circles are observational data 
	points for \mrk on March 19, 2001 from 
	\citet{fossati_2008:xray_tev}. Three grey dashed lines are also 
	plotted so that the readers can compare them to the trend of the 
	flux-flux relationship.
	Right: Cross-correlation between \xray and TeV \gray fluxes based on the 
	light curves, excluding the first 100\ ks. The confidence interval for the 
	lag based on the correlation function is marked with the horizontal solid 
	blue line, while the estimated lag is identified with the 
	dashed blue vertical line.}
	\label{fig:421_correl}
\end{figure*}

Since we have multiple cases, each with a similar set of analyses that would be
better understood when compared with each other, we merely list the findings of
each case and the variability analysis methods in this section, 
and leave the interpretation
of these findings to \S \ref{sec:discussion}.
The SEDs and light curves are not the focus of our study in this work.
The simulated observational SEDs and light curves for each case are shown here
to verify that we are using model parameters that are generally consistent with
what we know about \mrk.

\subsection{Case 1: Injection associated with acceleration}
\label{sec:injacc}

This is the benchmark simulation case for our study.
The SED and light curve samples are shown in Fig\,\ref{fig:421_300}. The conversion between apparent luminosity and flux is performed for 135 Mpc as the luminosity distance of \mrk.
Since the results are under the influence of stochastic fluctuations, we only
aim to approximately match the simulated SEDs with the observation. Likewise,
the simulated and observational light curves cannot be identical.
They are shown to demonstrate that they generally have similar shapes and
amplitudes.
The first 100 ks of the simulated light curves include the initial build up
of the photon field and the stabilization of the electron energy distribution.
We do not consider this period in subsequent analysis to avoid any unwanted 
effects from the initial condition.

\subsubsection{Power Spectral Density}
\label{sec:result:psd}

In our simulations, the total duration T of equation \ref{eq:psd} is 3000 ks minus the first 100 ks, 
which is considered the setup phase of the simulations. 
The PSD points are further binned into 20 logarithmically evenly spaced 
frequency bins from $8.8\times10^{-7}$ Hz to $5.9\times10^{-4}$ Hz. The first
two points are exempt from binning because of the sparseness of points 
at those frequencies,
making 22 the total number of final PSD points for each energy.

The resulting PSD for the entire simulation
(excluding the first 100 ks) is shown in Fig.\ref{fig:421_full} together with 
the full light curves. 
The PSDs therefore show that
the high-energy \xray flux is most variable within the synchrotron component.
Within the inverse Compton (IC) component at higher energy, 
the GeV \gray and TeV \gray have similar level of variability.

\begin{table}
	\tabcolsep 2pt
	\centering
	\caption{ Relaxation time $\tau_\mathrm{relax}$ in ks for various energy bands, 
	obtained by fitting the PSDs with equation \ref{eq:ou_psd}. 
	}
	\label{tab:psd}
	\begin{tabular}{ *{7}{c}}
	\hline
	Energy & Case 1 & Case 2
	& Case 3  \\\hline
	1-3 eV & 115 & 54 &70 \\[2pt]
	0.5-50 GeV & 64 &36 &68 \\[2pt]
	0.4-10 TeV & 27 &24 &52 \\\hline
	2-4 keV & 21 &23 &48 \\[2pt]
	9-15 keV & 14 &18 &28 \\[2pt]
	20-40 keV & 12 &15 &20 \\\hline
\end{tabular}
\end{table}

The PSDs follow a broken power-law distribution, which is expected from O-U processes.
However, the non-smoothness of the PSDs at low frequency indicates 
considerable uncertainty. This might hamper our identification of the breaks.
In order to increase the reliability of the PSDs, we repeated the simulation leading to
Fig.\,\ref{fig:421_full} with a
different sequence of random numbers. The PSD of this separate
simulation is shown in the left panel of Fig.\,\ref{fig:421_combined}. An average of
the PSDs from the two simulations is shown in the right panel of Fig.\,\ref{fig:421_combined}.

Apart from the red/white noise above/below the break frequency, the PSDs also flatten to white noise at the highest frequencies.
This is a natural result of the uncertainty in
the Monte Carlo method used for the radiative transfer, which renders 
unreliable variability below certain threshold.
This is also a great example that shows how the PSD can be used to
identify at which time scales and amplitudes the variability in a light curve 
should be suspected of containing mostly noise.

We fit the average PSDs of the two simulations with equation \ref{eq:ou_psd}.
Only frequencies below $10^{-4}$~Hz are considered in the fitting,
in order to avoid the white noise at low variation power caused by Monte-Carlo
fluctuation. The best-fit relaxation times, $\tau_\mathrm{relax}$, are listed in Table \ref{tab:psd}.
We can see that for both the synchrotron and IC emission, $\tau_\mathrm{relax}$ becomes 
smaller with increasing photon energy. For the X-ray synchrotron 
bands, $\tau_\mathrm{relax}$ is close to both the acceleration decay time, 
$\tau'_\mathrm{decay}/\Gamma=20$~ks, and the light crossing time, $\tau'_\mathrm{cross}/\Gamma=10$~ks.

\begin{figure*}
	\includegraphics[width=0.33\linewidth]{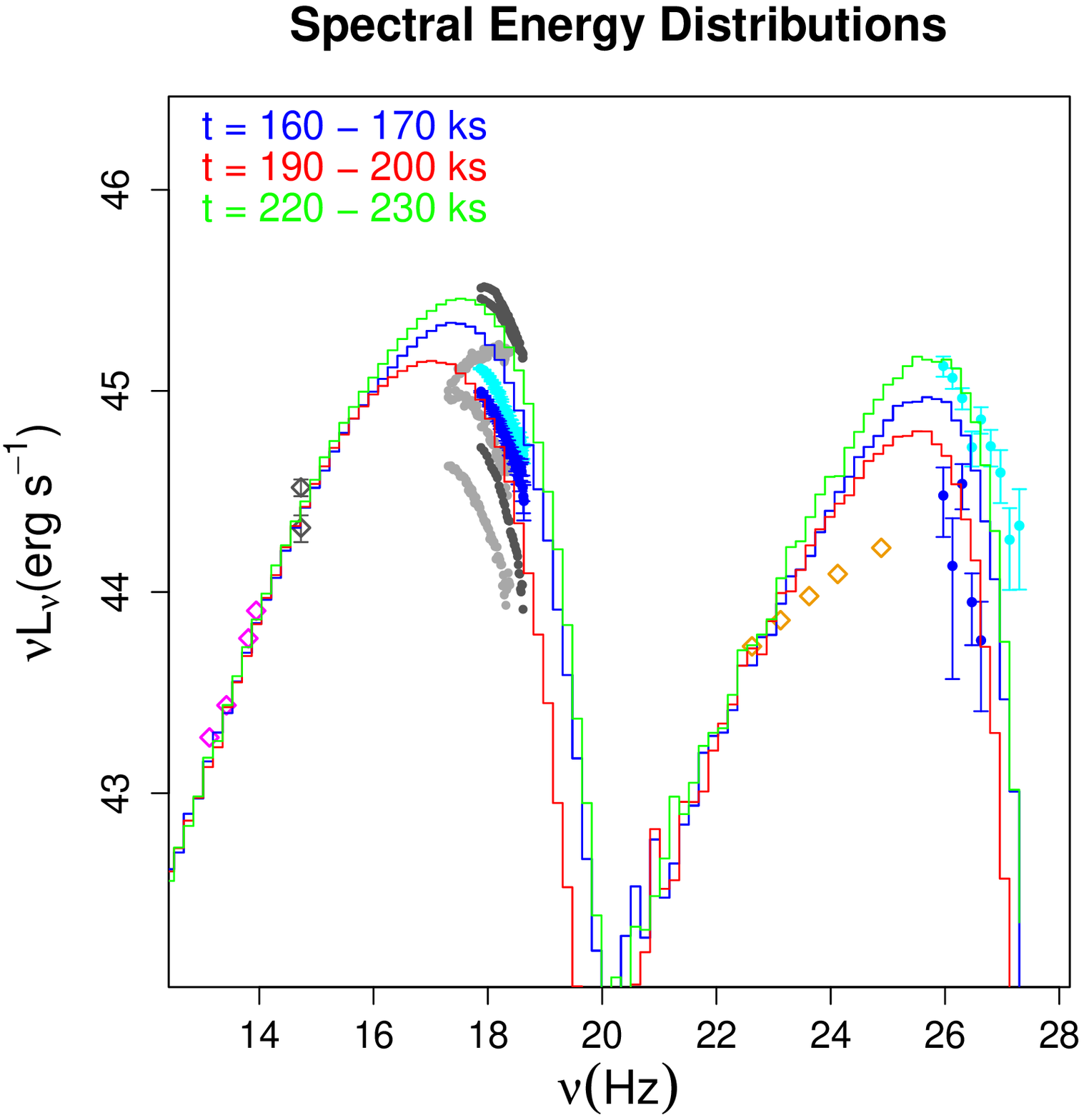}
	\includegraphics[width=0.33\linewidth]{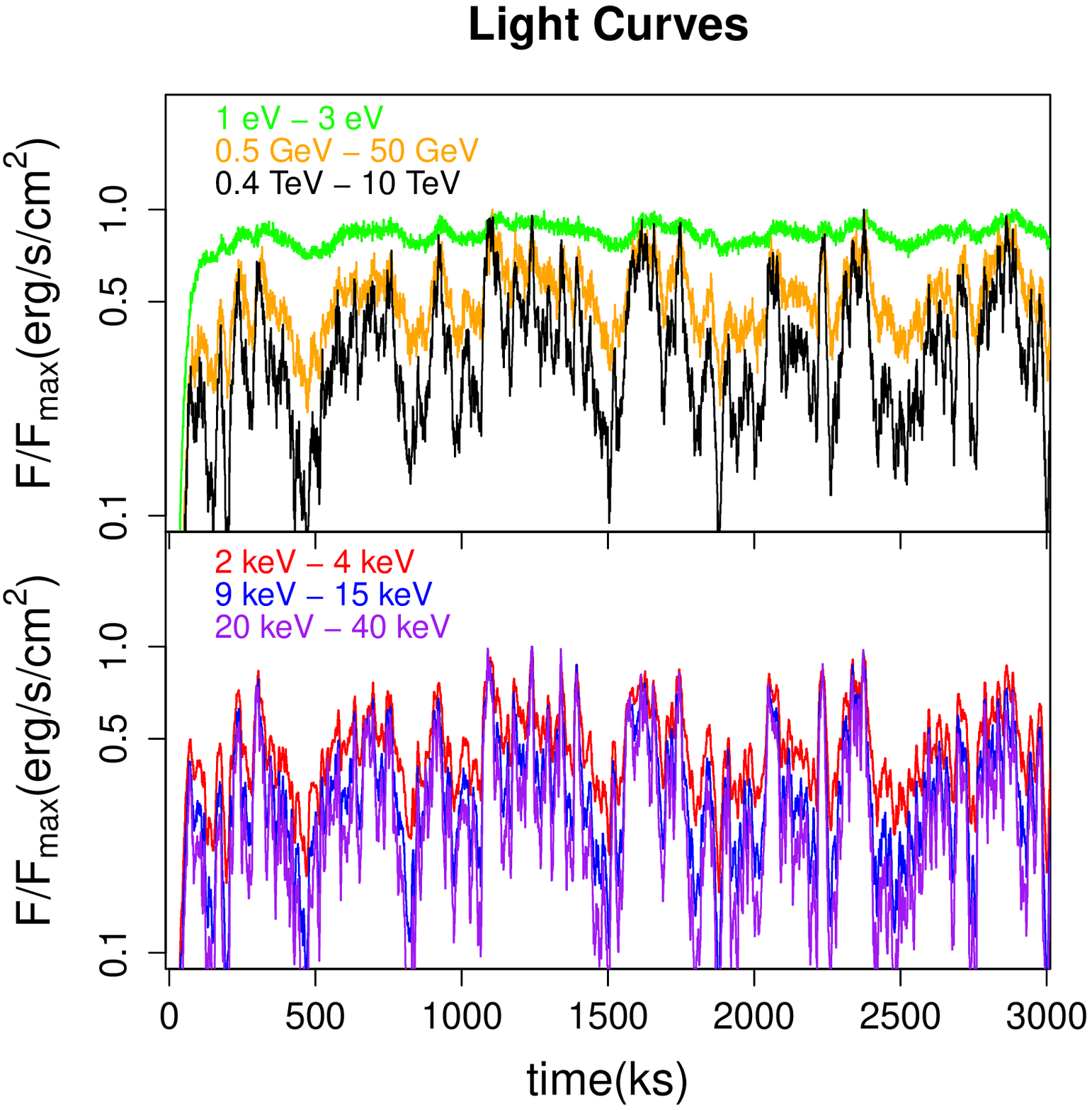}
	\includegraphics[width=0.33\linewidth]{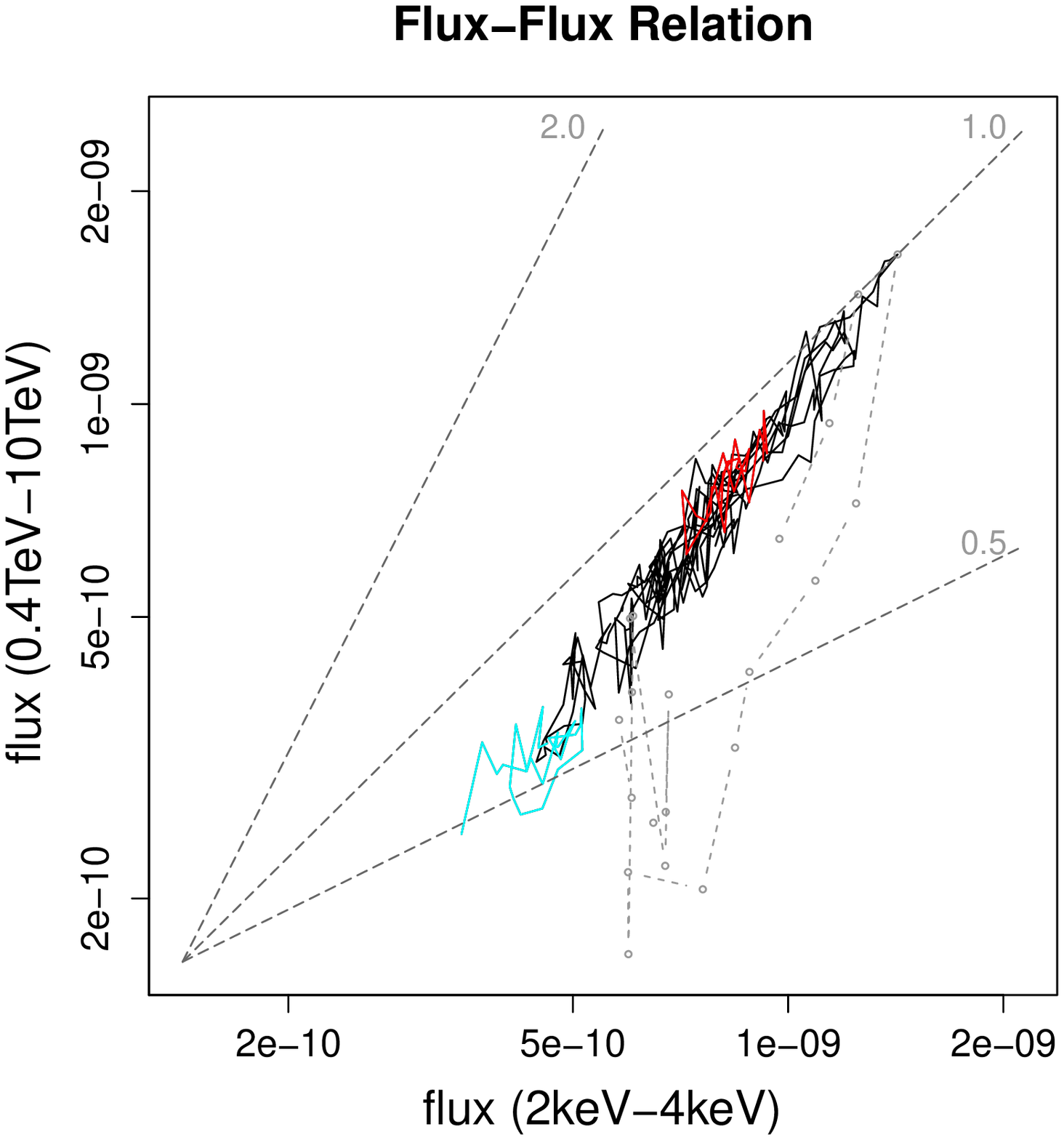}
	\includegraphics[width=0.33\linewidth]{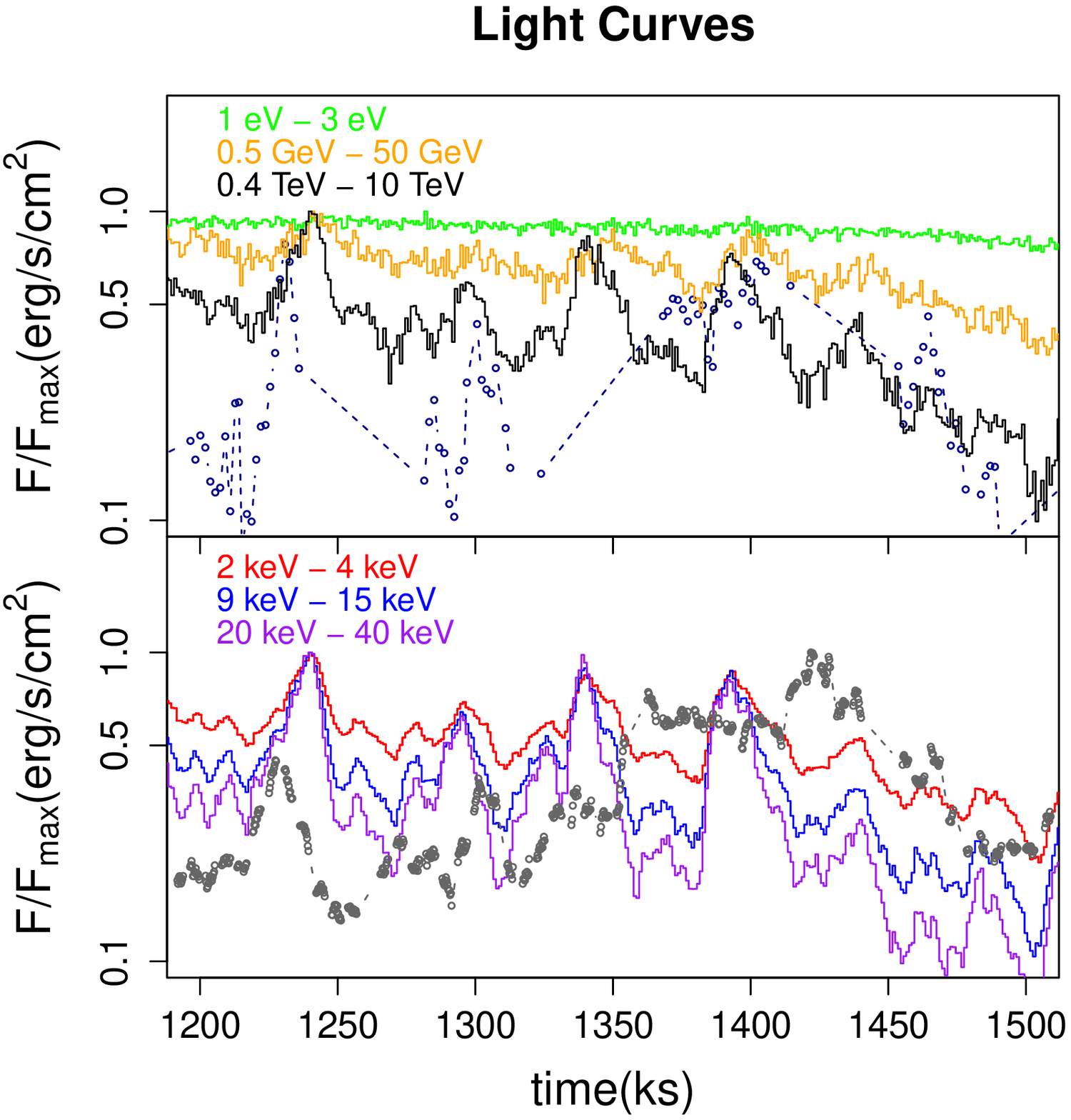}
	\includegraphics[width=0.33\linewidth]{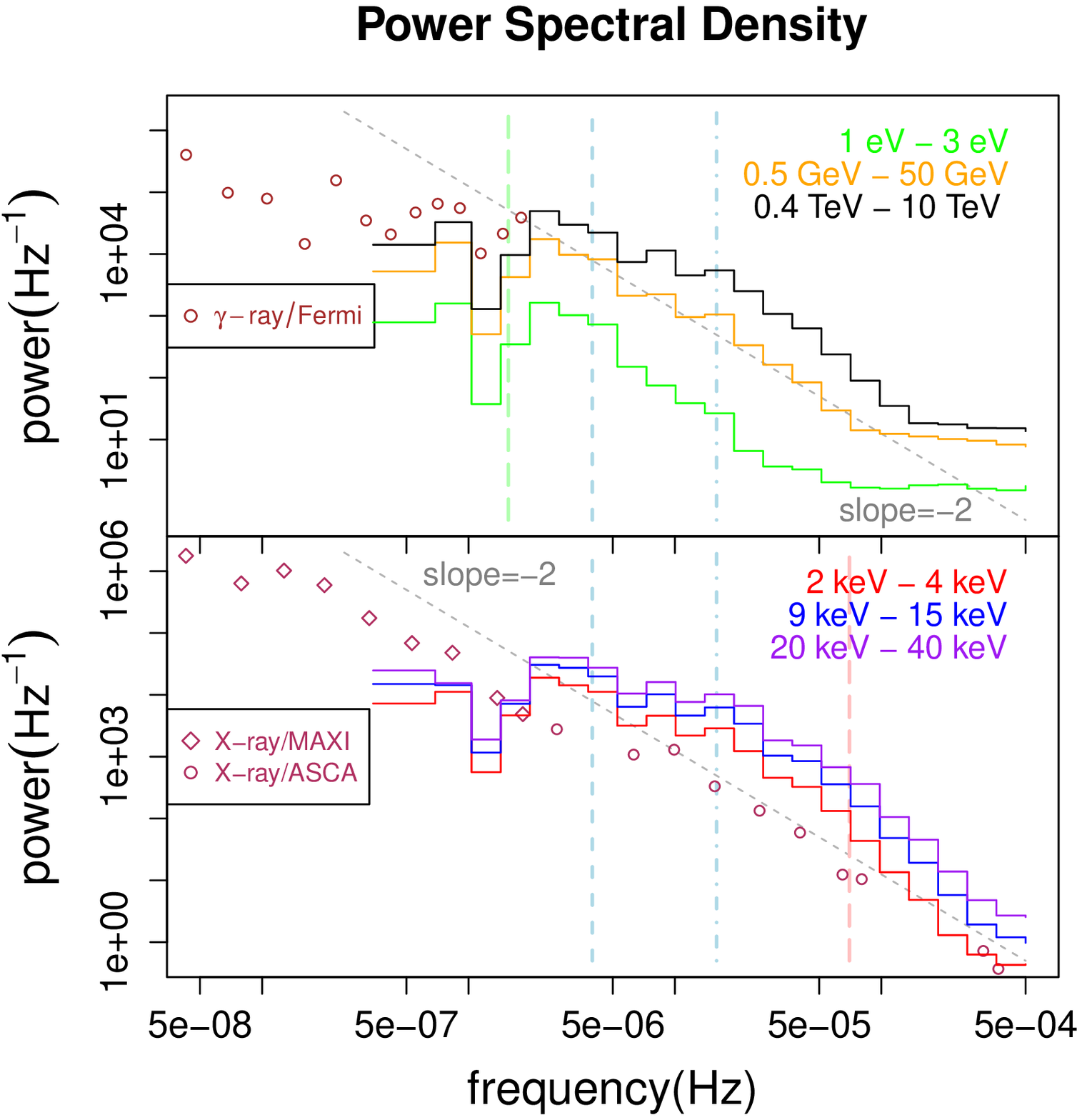}
	\includegraphics[width=0.33\linewidth]{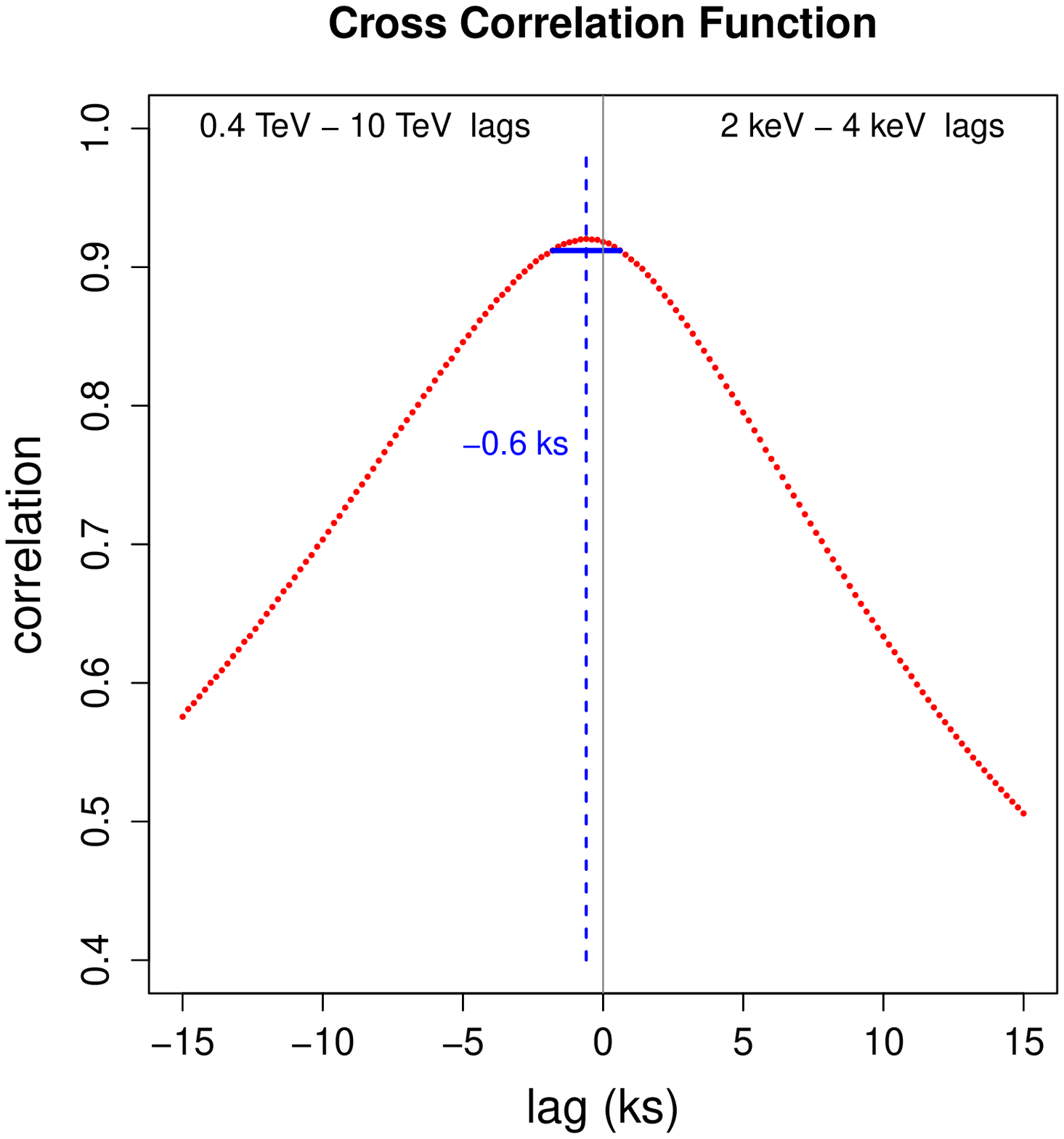}
	\caption{Figures similar to those shown in Fig. \ref{fig:421_300}, 
	  \ref{fig:421_full} \& \ref{fig:421_correl}, 
	  but for Case 2 with fixed particle injection 
	rate (\S\,\ref{sec:cpick}). 
	Figures include SEDs, full light curves, flux-flux correlation 
	in the top panel, as well as short light curves, PSDs, cross-correlation in the 
	bottom panel.}
	\label{fig:cpick}
\end{figure*}
\subsubsection{\xray/\gray Correlation}

We show the flux-flux amplitude correlation and the cross-correlation function
for the benchmark case in Fig.\,\ref{fig:421_correl}.

In order to obtain the uncertainty of the cross-correlation, 
we resample 
one of the two light curves without replacement, \ie the time series of
the light curve is reshuffled randomly. Then we calculate the correlation value of this reshuffled light curve with the other light curve, with zero lag.
Since now the two light curves should be completely uncorrelated, the expected
correlation is zero. Any non-zero values are assumed to be caused by random
fluctuations around zero.
By repeating this process for 2000 times, we get 2000 different 
cross-correlation values. The standard deviation, $\sigma_\mathrm{cc}$, for those 2000 values is taken as the uncertainty of the correlation.
We use the maximum of the correlation in Fig.\,\ref{fig:421_correl} minus 
$\sigma_\mathrm{cc}$ to set the confidence interval for the lag based on the correlation function,
while the estimated time lag is the mid-point of this interval.
For the purpose of maximizing the precision of the lags detected, we use 
0.2-ks time bins instead of the 1-ks bins used in the other plots.
This choice still makes sure that at all wavebands the number of Monte Carlo 
photons in most bins is larger than 100.

In this case the amplitudes of \xray and TeV flux show a quadratic relationship
very similar to the observed relationship in \mrk. The cross correlation
suggests that TeV \gray variability lags that in \xrays by 2.1 ks.

\subsection{Case 2: Fixed injection rate}
\label{sec:cpick}
In this case particle injection is constant in the acceleration region along the spine of the cylinder. Plots in format similar to those shown in 
Fig.\,\ref{fig:421_300},\ref{fig:421_full} \&\ \ref{fig:421_correl}
are shown in Fig.\,\ref{fig:cpick}. The fitted parameters for the PSDs
are shown in Table\,\ref{tab:psd}.
We can see that the PSDs in this case show similar trends as the ones observed
in Case 1, but compared to that case,
the \xray PSDs here generally break at longer time scales, or lower frequency,
while the optical and \gray PSDs break at shorter time scales, or higher frequency.
Another difference is that TeV \gray lightcurve shows a significantly higher fractional 
variability than that of GeV \grays.
The flux-flux correlation is shown to be close to linear rather than quadratic.
The cross-correlation function shows a small TeV \gray lag of 0.6 ks,
that is statistically compatible with zero lag.

\begin{figure*}
	\includegraphics[width=0.33\linewidth]{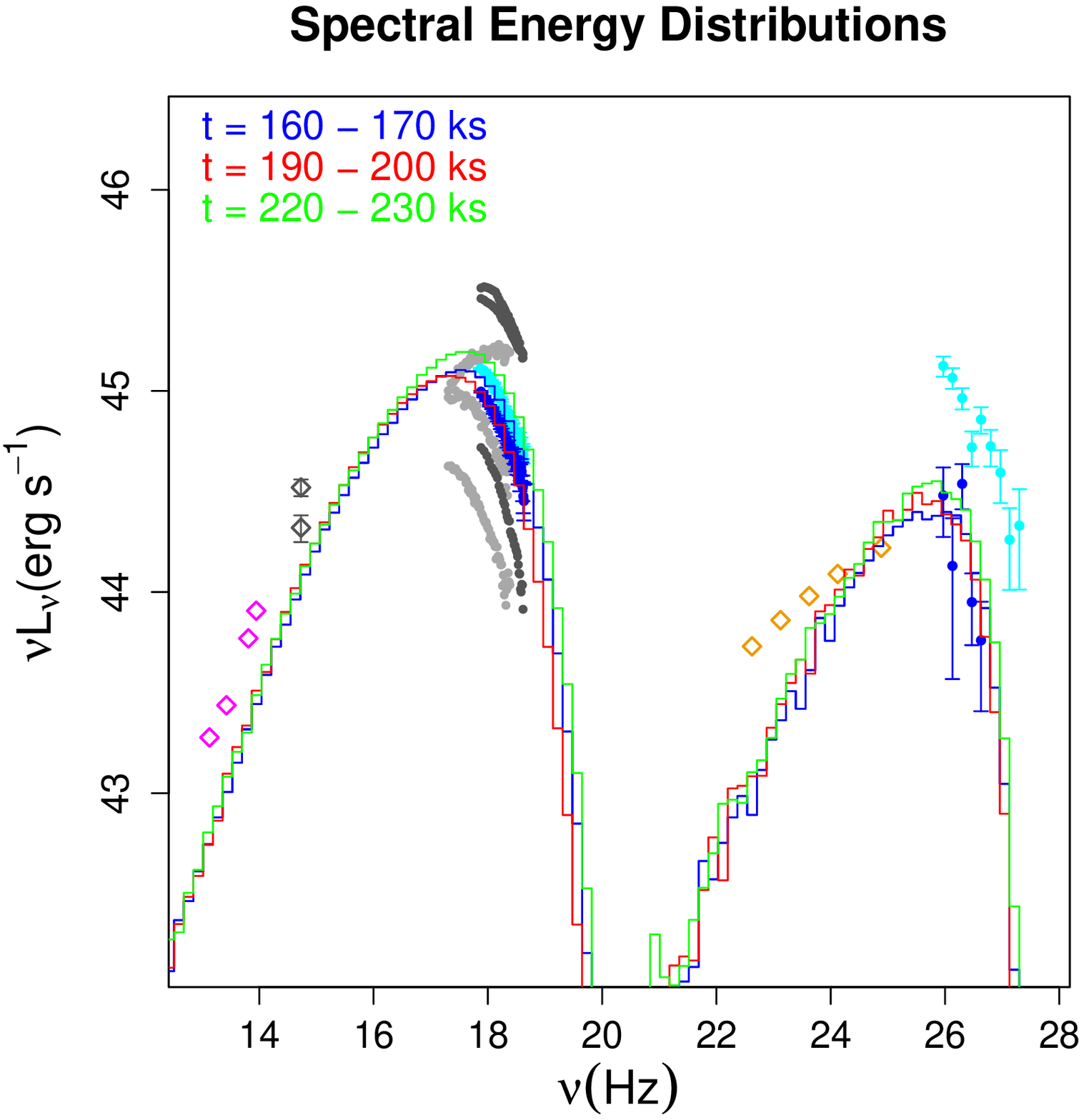}
	\includegraphics[width=0.33\linewidth]{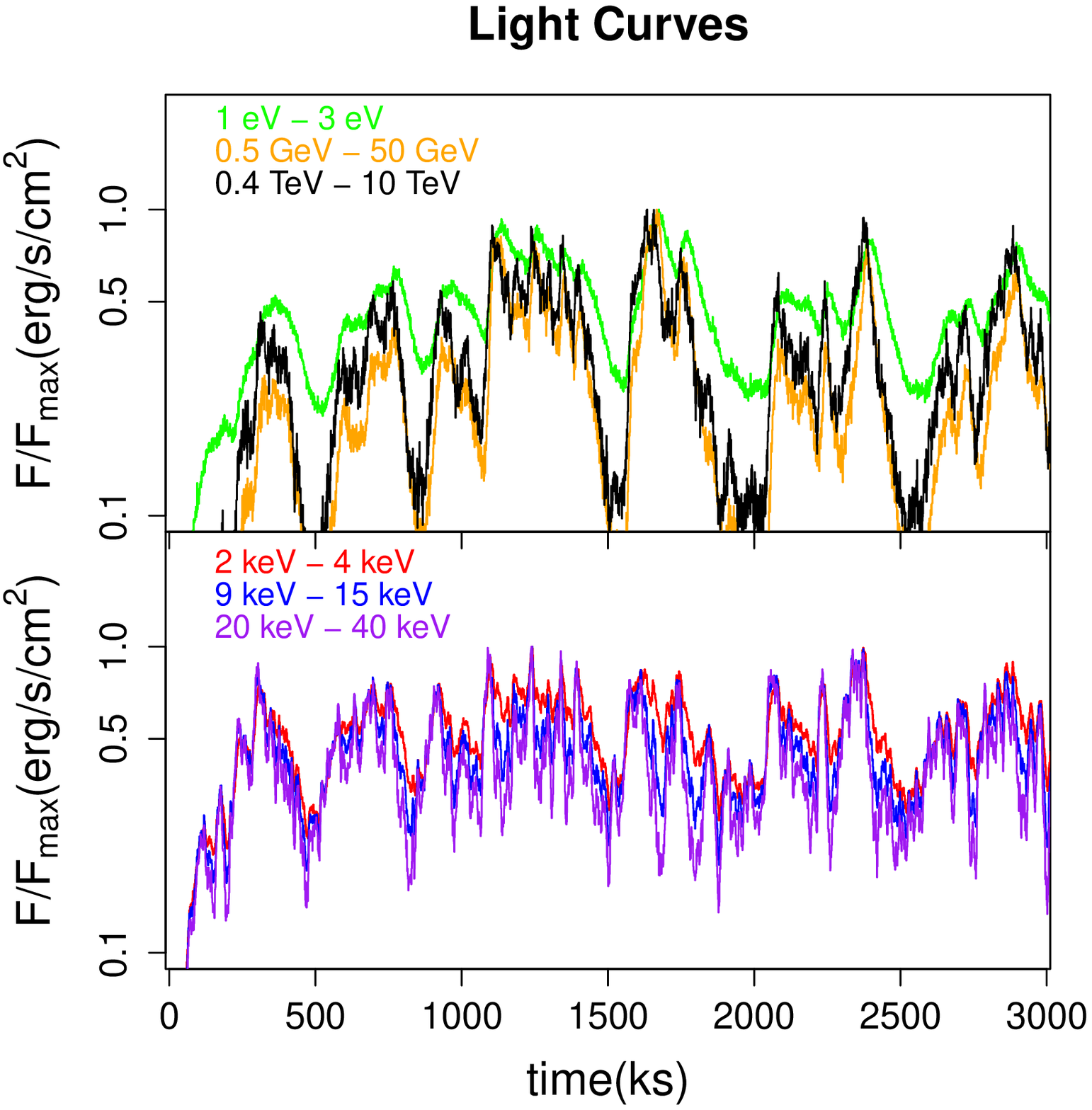}
	\includegraphics[width=0.33\linewidth]{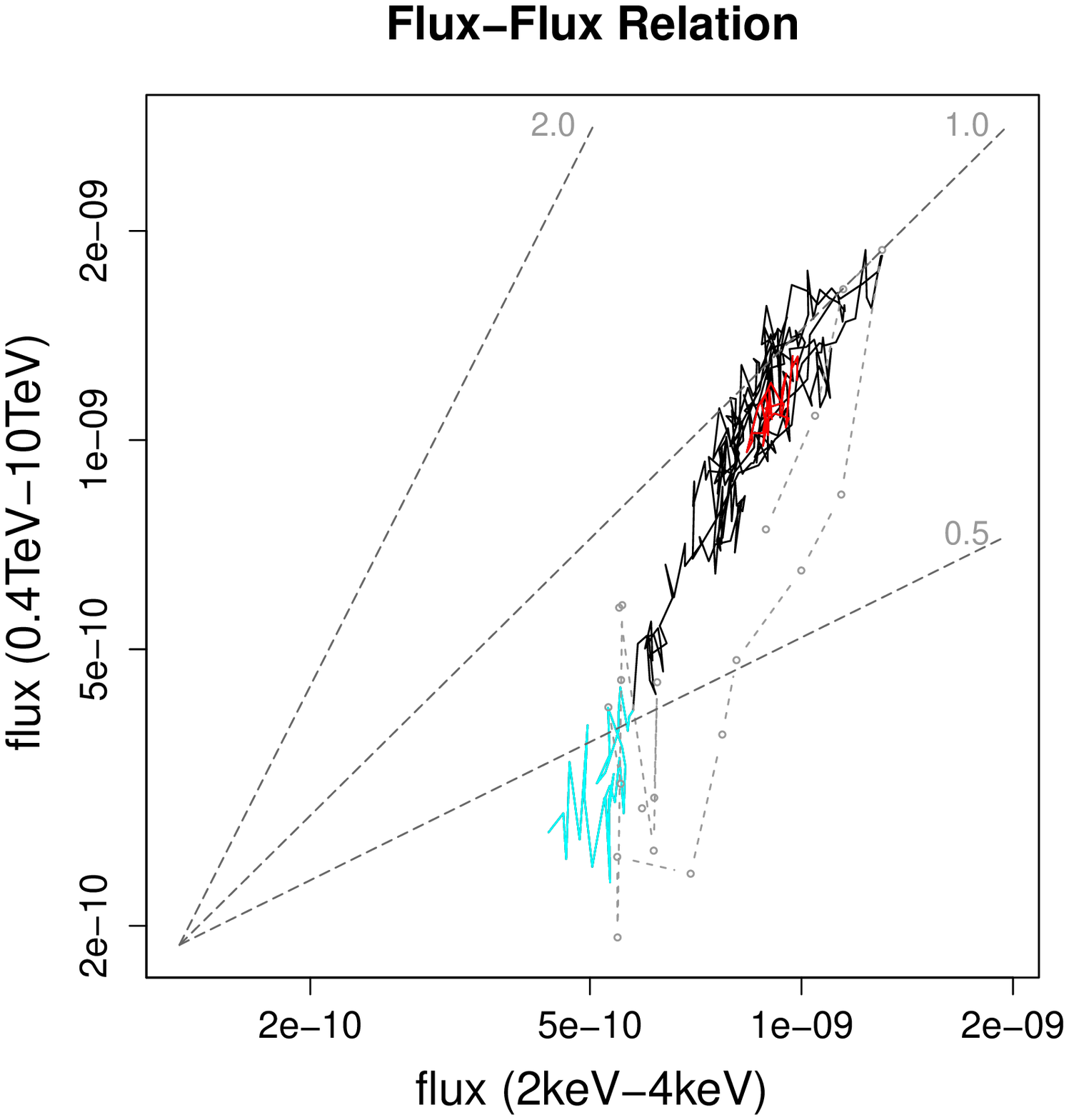}
	\includegraphics[width=0.33\linewidth]{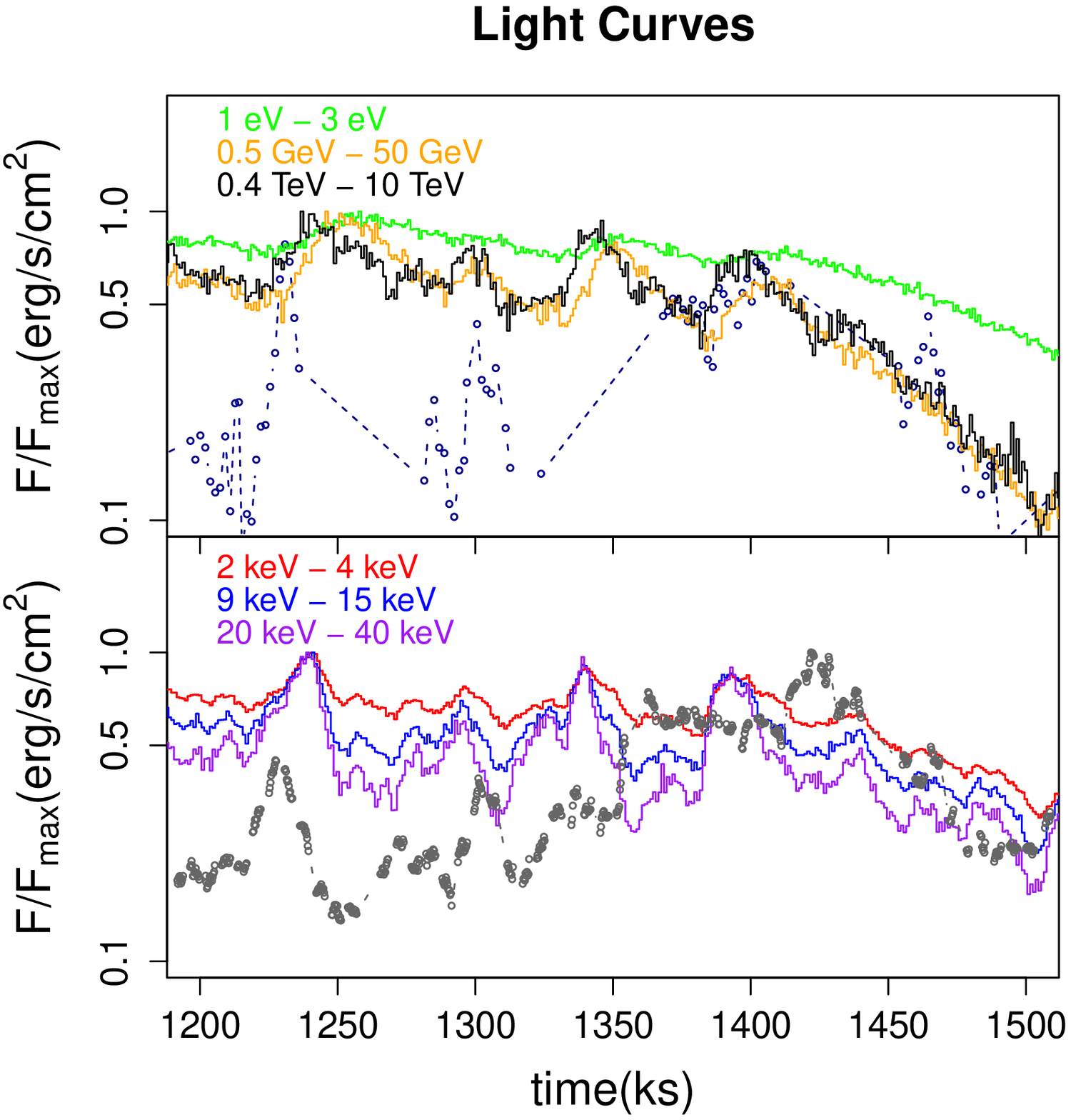}
	\includegraphics[width=0.33\linewidth]{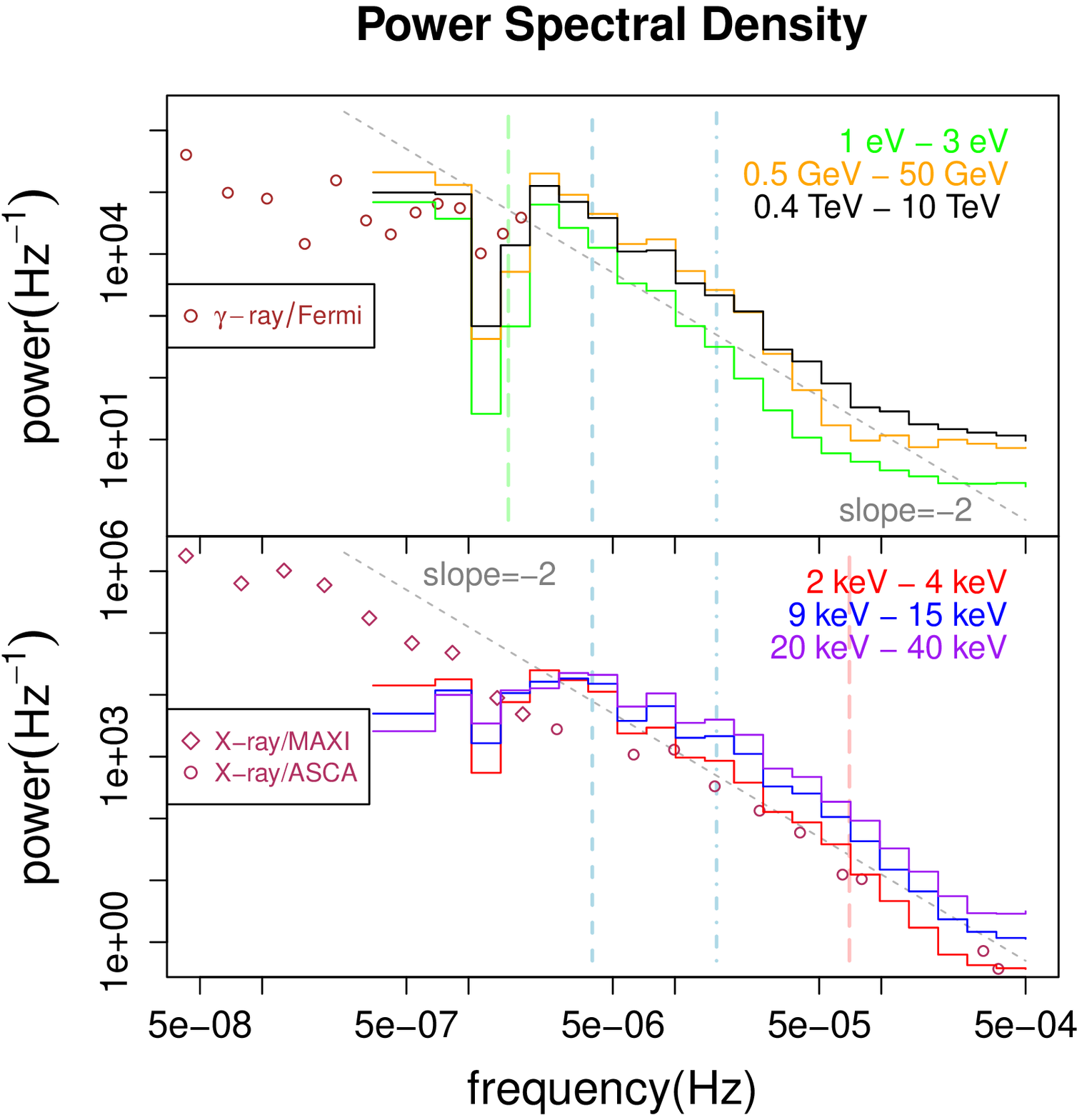}
	\includegraphics[width=0.33\linewidth]{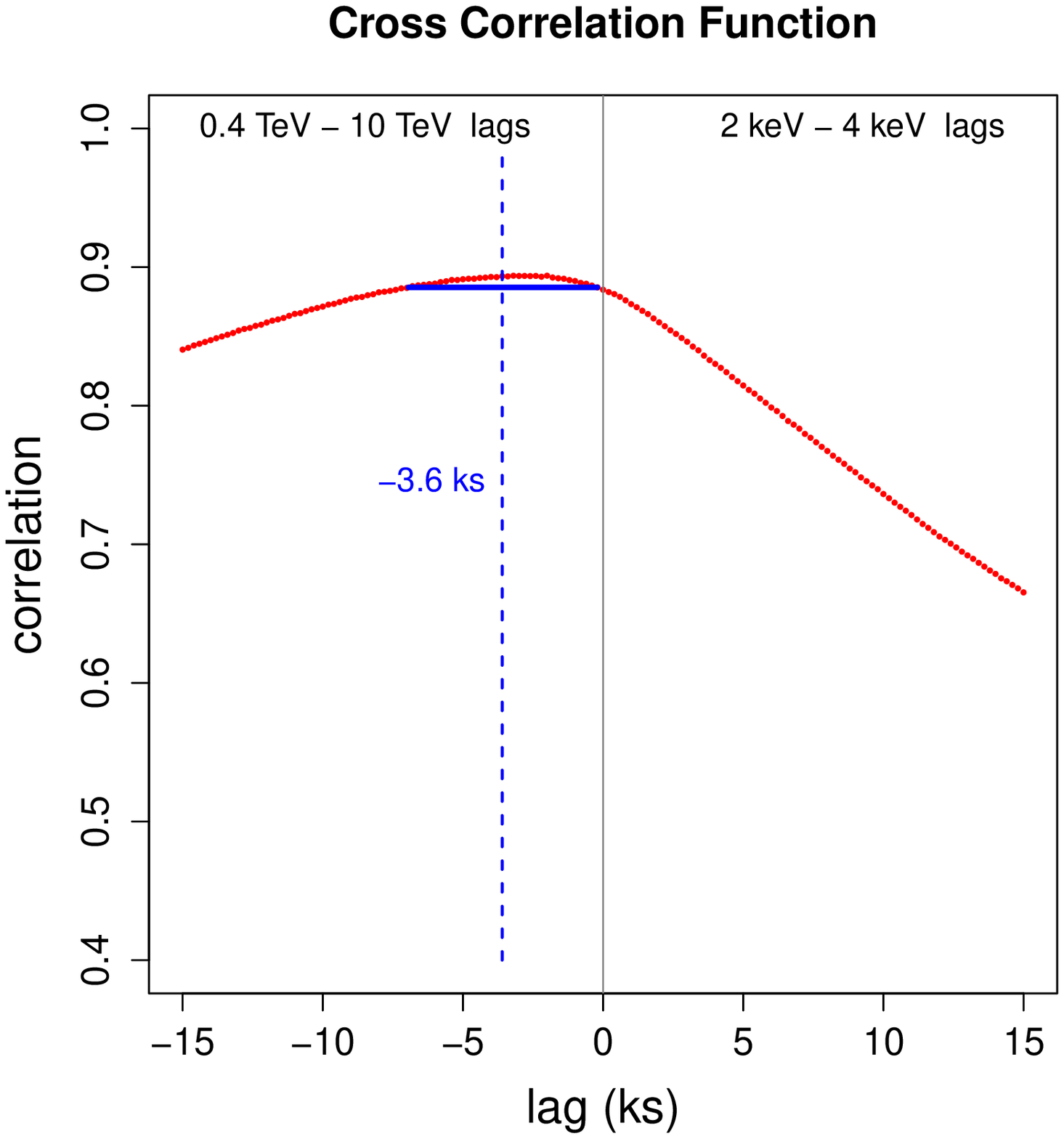}
	\caption{Figures similar to those shown in Fig. \ref{fig:421_300}, 
	  \ref{fig:421_full} \& \ref{fig:421_correl}, 
	  but for Case 3 with slower acceleration decay
	(\S\,\ref{sec:l80}). 
	Figures include SEDs, full light curves, flux-flux correlation 
	in the top panel, as well as short light curves, PSDs, cross-correlation in the 
	bottom panel.}
	\label{fig:l80}
\end{figure*}

\subsection{Case 3: Slower acceleration decay}
\label{sec:l80}
Compared to Case 1, here the acceleration decay time scale, 
$\tau'_\mathrm{decay}$, is changed from $2Z'/c$ to $4Z'/c$.
The results are shown in Fig.\,\ref{fig:l80}. Mostly they are consistent with
the results of Case 1. The effect of the longer acceleration decay time is that
most of the fitted relaxation times $\tau_\mathrm{relax}$ (except optical) for the PSDs 
become roughly a factor of 2 larger, which is a result of
doubling of the acceleration decay time. The complication
with the optical PSD is that, based on the $\tau_\mathrm{relax}$ obtained in Case 1, 
the one expected from the current case should be about
230 ks, which puts $1/2\pi\tau_\mathrm{relax}\sim7\times10^{-7}$Hz close to the edge of the 
covered frequency. This means the frequency range we use might not be 
sufficient to accurately identify the PSD break of optical emission 
in this case.
Therefore the value obtained here should be treated with caution.

\section{Discussion}
\label{sec:discussion}

\subsection{Comparison with Observation}
\label{sec:comparison}

In all the simulated cases, there is a good match between the simulated
and observational SED,  except the spectral indices at \grays. The SED observed by \fermi appears to be softer, which is at least partially caused by the fact that the \fermi data points are 4-year averages, while the other points and the simulation are snapshot SEDs at flaring state. The SED at very high energy (VHE) \grays (\grays above 300 GeV) is also always slightly
softer than the observed one \citep[see also][]{fossati_2008:xray_tev}
\footnote{This issue will not be solved by the consideration of the 
 extragalactic light absorption \citep{finke_2010:ebl:712.238}
 because that effect will only further soften the simulated \gray spectrum.
}.
This softer \gray spectrum results in simulated VHE \gray light 
curves that are dominated by photon flux from the low-energy end 
at around 500 GeV, which should vary much slower than flux above 1 TeV energy. Therefore the simulated VHE \gray light curves do not
fully reproduce certain very fast variations as seen in 
observations.
In general, the reason that observational \xray and \gray light curves 
are plotted together with their simulated counterparts is not for detailed fitting, 
but only to show that they exhibit variations of comparable amplitude and time scale.

The fractional rms variability shown in the simulated and observational PSDs 
are reasonably close.
However, the observational PSDs do not exhibit the 
interesting breaks we look for. 
The \xray PSD has a slope close to -2, consistent with a characteristic time larger than those covered here.
On the other hand, the PSD in \fermi \grays has a PSD slope close to 0, consistent with a high frequency break. 
It may appear attempting to explain the \gray light curve with fast decay and \xray light curve with slow decay because of the un-observed high/low frequency PSD breaks. But 
such fast decay in \grays typically implies a fractional rms variability larger than that in \xray. This is not the case according to those same observed PSDs.
This inconsistency between the \xray and \gray PSD slopes poses
 a challenge to both leptonic and hadronic emission models.
 However, it should be noted that even though the \fermi PSD already has the
white noise removed \citep{abdo_etal:2010:light_curves_and_variability}, 
the remaining variability amplitude for \mrk is similar to the white-noise amplitude. Considering that the white-noise level of the observation has its own uncertainty, one should view those data points with extra caution.

The lack of any break in the observational PSDs indicates that in reality the 
break may only exist at longer or shorter time scale, 
which is beyond our current observation capability.
The interpretation of the \xray PSD might be further complicated if 
the quiescent state of the \xray has a significant contribution from IC or hadronic emission, for which particles should have a long cooling time at the corresponding energy. This may result in a PSD with two components both having slope -2, but not aligned with each other.\footnote{Notice the \xray PSDs measured by \maxi 
and \asca \citep{isobe_2015:maxi_421_psd:798.27} are also not
perfectly aligned. But the misalignment is so small that it is probably an artifact from the different calibrations of two instruments rather than a result of two PSD components.}
Since the study of the PSD break and its relationship with various time scales
in the system is the primary goal of the current work, 
we only chose simulation parameters that lead to PSD break within the frequency range
we cover,
therefore the simulated PSD
is guaranteed not to match the observed PSD by design.
This mismatch does not compromise the validity of the model as a whole,
but only means we have to keep in mind that the relaxation time 
derived in this work is not the real system relaxation time for \mrk
, or what we see in observation is a mixture of multiple relaxation times instead of a single relaxation time.

As we can see, with our current model, we can now simulate stochastic variability in many different wavebands. But the observational data we can compare to, especially the PSDs, are still relatively sparse. Future observations may provide us with PSDs that span a larger range of time scales (higher cadence and longer total time \footnote{Notice that the observation does not necessarily need to be high-cadence all the time. High-cadence data for certain short periods of time would be sufficient for obtaining the PSD at high frequency.}), at more wavebands including the optical and VHE \grays. These observations will be essential in discriminating blazar models and enriching our knowledge of blazar variability.

\subsection{Correlations}
\label{sec:discuss:cor}
\citet{fossati_2008:xray_tev} have shown that in an SSC model of \mrk, seed photons with energy between 5~eV and 500~eV are most important for TeV-band emission. By considering the larger Doppler factor used in our cases, this range is further reduced to 2-200~eV. At the same time, the electrons
that are responsible for the IC emission are emitting synchrotron photons in \xray.
Therefore the TeV flux can be considered as representative to a product of 
the \xray flux and the lower energy flux that can be  
represented by the optical flux. This aspect also affects the phase correlation of TeV \grays
with the \xrays. The optical light curves for cases 
1 and 3 are not perfectly symmetric, with
the decay being generally slower. This is because the diffusion-caused 
escape from the entire emission region, even though faster than cooling
at these energy, is still relatively slow compared to the changes of injection.
It is this asymmetry that causes its product with the \xray to be
delayed compared to the \xray. This is likely the main cause of the marginal TeV lags
observed in those two cases.

Another subtle effect that alters \gray lightcurves is internal light-travel-time effects (LTTEs), \ie the effects of the extra time the seed
photons need to spend traveling within the emission region before they
are IC scattered by high-energy electrons. That extra time is typically
a fraction of the light-crossing time of the emission region, which is
roughly 10 ks in terms of delay in the observed signal in our current cases.
This 'fraction' turns out
to be $\sim0.4$ in a case where the high-energy electrons are homogeneously 
distributed and do not cool significantly 
\citep{chen_2014:bamplify:441.2188C}. 
In the current case where acceleration-diffusion causes 
the emission region to be dominated by its central portion, this fraction is likely much smaller, and so is the impact of internal LTTEs.

In Case 2 the optical flux is relatively stable compared to 
the other cases, due to the constant rate of particle injection at low energy. This is the main reason why in this case the flux-flux correlation is close to linear, while in the other two cases it is quadratic. 
In the study of \citet{chen_etal:2011:multizone_code_mrk421}, a quadratic relationship is seen in the rising phase of the flare, while a linear relationship is 
evident in the decay phase. That work neglected particle
escape which causes a decrease in the number of the low-energy electrons that emit
the seed photons, whereas in the new model spatial diffusion offers
a way for particles to escape once the injection rate decreases.

It is interesting to note that even though the observational data plotted for 
comparison show a quadratic relationship, there are also other times when a linear trend \citep{fossati_2008:xray_tev,
magic_veritas_2015:mrk421_2010:578.22}, and occasionally even a cubic correlation \citep{aharonian_etal:2009:pks2155}, is observed. It is also important to notice that the relationship 
is sensitive to the choice of \xray energy band
\citep{katarzynski_etal:2005:tev_x_correlation},
because bands at higher energy generally tend to be more variable,
as can be seen in the simulated \xray light curves and PSDs,
as well as in observations \citep{magic_veritas_2015:mrk421:576.126}.

Our simulations suggest that a quadratic relationship between \xray and 
TeV \gray is accompanied by TeV lags on the order of several ks,
while a linear relationship is associated with insignificant lag of less than 
1ks. Interestingly, \citet{fossati_2008:xray_tev} found both a quadratic
relationship between \xray and TeV \gray, and a TeV \gray / soft \xray lag 
of 2.1ks
during the flare in March 19, 2001, while seeing a linear relationship and no
significant lag during the other time periods of that same week.

Even though the TeV lags in \S\,\ref{sec:injacc} \& \ref{sec:l80} are more 
significant than the one in \S\,\ref{sec:cpick}, the uncertainty is still 
large in those cases. we therefore can not draw
further quantitative conclusions regarding the TeV lags.

\subsection{Power Spectral Density}
\label{sec:discuss:psd}
A key feature of the PSDs in an O-U process, or in our simulations, is a
break where the PSD changes from white noise at low frequencies to
red noise at high frequencies. Because the break frequency is defined as $1/(2\pi\tau_\mathrm{relax}$),
we can use it to infer the relaxation timescale of a system. For example, 
\citet{finke_2014:fourier:791.21,finke_2015:fourier_kn}
studied the electron-transport equation in the Fourier domain with an 
analytical approach and found the breaks to be associated with the
inverse of the cooling time. However, in order to make the equations
analytically tractable, they sacrificed considering
more complicated processes and effects such as the particle acceleration, 
the spatial inhomogeneity, and the LTTEs.
Instead, they used direct 
injection of high-energy electrons assuming they are accelerated 
instantaneously, leaving radiative cooling as the only process that acts 
on the electrons.

With our more complex acceleration-diffusion model, 
we find that the system relaxation
in blazars reflects more than just the cooling. For 
2-4 keV emission the cooling time of the underlying electrons is in fact
an order of magnitude smaller than the relaxation time, 
$\tau_\mathrm{relax}$, identified in
\S\,\ref{sec:injacc}. The connection between $\tau_\mathrm{relax}$ in the \xray band and
the timescale on which the acceleration rate changes 
($\tau'_\mathrm{decay}/\Gamma$)
is established in \S\,\ref{sec:l80}, where $\tau_\mathrm{relax}$ is found to double when 
$\tau'_\mathrm{decay}$ doubles.
This connection is explained as follows: Most of the time the electrons
are in an equilibrium between acceleration and cooling, at least locally in
the acceleration cells. Emission from these cells also dominates
the high energy end of the spectrum (see Paper I). 
So at those energies the emission
changes follow the acceleration changes. 
However, their relationship is not
linear, and so $\tau_\mathrm{relax}$ and $\tau'_\mathrm{decay}/\Gamma$ are not always the same.

The adherence of the flux to the acceleration is further enhanced by the assumed relation
between particle injection and acceleration (\S\,\ref{sec:injacc} \& \ref{sec:l80}).
Once this relation is broken 
(\S\,\ref{sec:cpick}), 
the flux variability becomes slightly weaker on short timescales,
and values of $\tau$ are shown to increase.

The effects of cooling on $\tau_\mathrm{relax}$ are evident from the fact 
that $\tau_\mathrm{relax}$ is energy dependent,
with $\tau_\mathrm{relax}$ being larger for lower energy in both synchrotron and 
IC component. This trend is caused by the fact that cooling cannot
instantaneously return
the system to a local equilibrium,
especially considering that there is counter-reaction from acceleration
even when it is relatively weak.
The slower the cooling is at lower energy, the stronger 
it modulates the light curves
that are still mainly affected by $\tau'_\mathrm{decay}$.
It is also interesting to note that in the 1-3 eV optical band, 
  the lowest energy frequency
  for which we analyzed the PSD, the breaks are also consistent with being 
  caused by
  the synchrotron cooling time scale at that frequency.

Fitting breaks in \gray PSDs generally yields larger timescales than found in  
\xray PSDs.
For \fermi \grays, this is at least partly caused by the fact that they
originate from lower energy particles, which has longer cooling time.
For VHE \grays, another important reason might be that these
\gray emission reflects both the optical flux
and the \xray output. Therefore \gray variability may be better described as a 
mixture of more than one O-U processes, as proposed by \citet{sobolewska_2014:psd_mix_ou:786.143}.
Since the optical PSD breaks at low frequency, the \gray PSDs also begin to 
turn over at relatively low frequency, even though they do not necessarily
change to red noise as in a single O-U process.
In Case 2, the variability in optical flux is minimal,
therefore it only slightly alters the \gray PSDs, leaving the difference
in the breaks identified from the \gray and \xray PSDs relatively small.

\subsection{Fractional Variability}
\label{sec:discuss:var}
The fractional rms variability as represented by the PSDs show
that the amplitude of \xray variability increases with photon energy, consistent
with the observations of \mrk \citep{fossati_etal:2000:mkn421_temporal,
  fermi_2011:mrk421:736.131,
magic_veritas_2015:mrk421:576.126}.
However, the variability of the lower energy bands in both the synchrotron
and IC components depends on the model assumptions.
In the constant particle-injection case (Case 2), both optical and GeV 
\gray variability are weak, even though the \xray and TeV \gray are still
very much variable. This type of activity is observed in \mrk
\citep{fermi_2011:mrk421:736.131, chen_2014:op_421:349.909}
especially during its non-flaring state 
\citep{magic_veritas_2015:mrk421:576.126}.
If the injection varies a lot during flares (\S\,\ref{sec:injacc} \& 
\ref{sec:l80}), the variability at lower energies of both synchrotron and IC components
is increased due to the large variation in lower-energy particles.
In those cases the compatibility between the model and the observations is
no longer immediately apparent. 
But if one assumes that both situations (injection varies or not) occasionally happen 
in reality, one can notice that in real observations 
the cases where quadratic relationship between
\xray and TeV \gray exists, which correspond to the cases with larger optical or GeV \gray variability in our model, 
are not so common \citep{fossati_2008:xray_tev}. 
The rare occurrence of large radio-optical and GeV flares 
\citep[See][for a large GeV \gray and radio flare of \mrk in 2012]{
hovatta_2015:mrk421_2012flare:448.3121}
implies that they have little
contribution to the total fractional variability,
therefore it is not surprising that the overall variability in those bands 
is still lower than that in \xrays and TeV \grays.
Another possible explanation is that
at low energies the emission may have a
significant contribution from separate non-varying emission regions 
\citep{chen_etal:2011:multizone_code_mrk421}.

\section{Conclusion}
\label{sec:conclusion}

In this work we modelled the stochastic variation of blazar emission in many wavebands, using physically motivated acceleration and decay processes. The main findings include:
  
\begin{enumerate}
 \item Our simulations predict the quadratic relationship between \xray and 
   TeV \gray to be associated with TeV \gray lags with respect to \xray. 
   These quadratic relationship and lags are also accompanied by strong 
   optical and GeV \gray flares. 

 \item All those three features are results of large increase in 
   the injection into the particle acceleration process.

 \item Possible detections of PSD breaks in blazars will reveal the 
  characteristic relaxation time in blazars.
  An example of such relaxation is the decay of the particle 
  acceleration as modelled in our simulations.
  The breaks may be energy 
  dependent because of cooling, but they do not necessarily correspond to 
  the cooling time.

 \item The mismatch between the currently observed PSD slopes
   in \xrays and \grays presents a challenge to 
   existing blazar emission models.

\end{enumerate}

Our findings also call for better characterization of the statistical property of blazar variability in different wavebands, hence encourage future high-cadence and long-term monitoring campaigns.

\section*{Acknowledgements}

We thank the anonymous referee for comments that helped to improve the presentation of the paper.
We also thank Justin Finke for valuable discussions and comments.
XC and MP acknowledge support by the Helmholtz Alliance for Astroparticle Physics HAP funded by the Initiative and Networking Fund of the Helmholtz Association. MB acknowledges support from the South African Department of Science and Technology through the National Research Foundation under NRF SARChI Chair grant No. 64789.

\bibliography{refs_all}

\end{document}